\documentclass[amsmath,prd,showpacs,preprintnumbers,nofootinbib,superscriptaddress,secnumarabic, graphics,floatfix,nofootinbib,
tightenlines,nobibnotes,aps]{revtex4-1}
\usepackage{epsfig}
\usepackage{epstopdf}
\usepackage{amsmath}
\usepackage{amssymb}
\usepackage{graphicx}
\usepackage{caption}
\usepackage{color}
\usepackage{subcaption}

\def\be{\begin{equation}}
\def\ee{\end{equation}}
\def\ba{\begin{eqnarray}}
\def\ea{\end{eqnarray}}
\def\gz{g_{\zeta}}
\def\fnl{f_{\rm NL}}
\def\fnliso{f^{\rm iso}_{\rm NL}}

\def\gnl{g_{\rm NL}}

\def\taunl{\tau_{\rm NL}}

\def\nn{\hat{n}}

\def\kn{\hat{k}}

\begin{document}

\title{Scale and shape dependent non-Gaussianity in the presence of inflationary vector fields }

\author{Juan P. Beltr\'an Almeida}\email{juanpbeltran@uan.edu.co} 
\affiliation{Departamento de F\'isica, Universidad Antonio Nari\~no, \\ Cra 3 Este \# 47A-15, Bogot\'a D.C. 110231, Colombia}
\author{Yeinzon Rodr\'iguez}\email{yeinzon.rodriguez@uan.edu.co}
\affiliation{Centro de Investigaciones en Ciencias B\'asicas y Aplicadas,Universidad Antonio Nari\~no, \\ Cra 3 Este \# 47A-15, Bogot\'a D.C. 110231, Colombia}
\affiliation{Escuela de F\'isica, Universidad Industrial de Santander, \\ Ciudad Universitaria, Bucaramanga 680002, Colombia}
\author{C\'esar A. Valenzuela-Toledo}\email{cesar.valenzuela@correounivalle.edu.co}
\affiliation{Departamento de F\'isica, Universidad del Valle,\\  Ciudad Universitaria Mel\'endez,
 Santiago de Cali 760032, Colombia}



\begin{abstract}
We consider cosmological inflationary models in which vector fields 
play some role in the generation of the primordial curvature perturbation $\zeta$.
Such models are interesting because the involved vector fields naturally seed statistical anisotropy in the primordial fluctuations which could eventually leave a measurable imprint on the cosmic microwave background fluctuations. In this article, we estimate the scale and shape dependent effects on the non-Gaussianity (NG) parameters due to the scale dependent statistical anisotropy in the distribution of the fluctuations. For concreteness, we use a power spectrum (PS) of the fluctuations of the quadrupolar form: $P_\zeta(\vec{k})\equiv P_\zeta(k)\left[1+g_\zeta(k)(\hat{n} \cdot \hat{k} )^2 \right]$, where $g_{\zeta}(k)$ is the only quantity which parametrizes the level of statistical anisotropy and $\hat{n}$ is a unitary vector which points towards the preferred direction.  Then, we evaluate the contribution of the running of $g_{\zeta}(k)$ on the NG parameters by means of the $\delta N$ formalism. We focus specifically on the details for the  $f_{\rm NL}$ NG parameter, associated with the bispectrum $B_\zeta$, but the structure of higher order NG parameters is straightforward to generalize. Although the level of statistical anisotropy in the PS is severely constrained by recent observations, the importance of statistical anisotropy signals in higher order correlators remains to be determined, this being the main task that we address here. The precise measurement of the shape and scale dependence (or running) of statistical parameters such as the NG parameters and the statistical anisotropy level could provide relevant elements for model building and for the determination of the presence (or nonpresence) of inflationary vector fields and their role in the inflationary mechanism.  
\end{abstract}


\maketitle

\section{Introduction}\label{intro}
The study of statistical characteristics of the inflationary fluctuations such as non-Gaussianity (NG) and statistical anisotropy, and their signatures in the cosmic microwave background (CMB) and in the large scale structure (LSS), is a subject of major interest in modern cosmology 
\cite{Huterer:2010en}.  The interest in these subjects is justified by the fact that non-Gaussianity and statistical anisotropy signatures in the probability distribution of the CMB temperature anisotropies are sensitive to the specific details of the mechanisms ruling the dynamics of the early Universe;
thus, their  precise evaluation could provide relevant criteria to discriminate among the many models proposed to explain the origin of the LSS distribution that we observe today. Moreover, the detection of a significant signal of NG and statistical anisotropy would rule out the simplest models of inflation based on a single ``{\it slowly rolling}'' scalar field driving the inflationary mechanism. A possible way to generate significant levels of statistical anisotropy and NG is by introducing vector fields during the inflationary epoch as a  source of the primordial curvature perturbation $\zeta$. For this reason, inflationary models with vector fields have been studied with great interest during recent years, and the literature addressing the main features of these models is rich; see, for instance, Refs. \cite{Dimopoulos06, Ackerman07, Golovnev08, Dimopoulos08, Yokoyama08, Himmetoglu08, Karciauskas08, Dimopoulos09a, Dimopoulos09vu, Watanabe:2009ct, Watanabe10, Maleknejad11, Maleknejad:2011jw, Maleknejad:2011jr, Bartolo09a, Bartolo09b, Valenzuela09a, Valenzuela09b, Gumrukcuoglu10, Dulaney10, Dimopoulos11, Bartolo11, Bartolo12, Beltran11, Rodriguez13, Gomez:2013xza, Lyth13, Shiraishi13, Namba:2013kia, Adshead:2012kp, Adshead:2013nka, Ohashi:2013qba, Valenzuela11,Namba12gg,Abolhasani13,Fujita:2013qxa,Maeda:2013daa} (for reviews see Refs. \cite{Dimastrogiovanni10, Soda12, Maleknejad12}). Besides, inflationary models with vector fields are interesting on their own because their phenomenology is rich and also
because the statistical parameters, modified by the privileged directions inherent to the nature of the vector fields,
could eventually be tested 
with the required amount of precision. 

The main purpose of this paper is the evaluation and the detailed quantitative description of the principal features of the statistical parameters related to NG and statistical anisotropy in the presence of vector fields and the relations among them. 
We focus mainly on the first level of non-Gaussianity $f_{\rm NL}$ obtained from the three-point correlation function of $\zeta$, or equivalently, its Fourier transform in momentum space, the {\it bispectrum} (BS) $B_\zeta(\vec{ k}_{1}, \vec{ k}_{2}, \vec{ k}_{3})$. The bispectrum is sensitive to the configuration adopted by the wave vectors $(\vec{ k}_{1}, \vec{ k}_{2}, \vec{ k}_{3})$ in momentum space. This sensitivity  is exacerbated in the presence of vector fields  given that, in this case, the bispectrum not only depends on the magnitude of the wave vectors but also on their orientations with respect to a preferred direction induced by the vector field. There are many ways to parametrize the presence of statistical anisotropy in the correlation functions, but for simplicity and definiteness of the analysis, we restrict ourselves here to a parametrization generated from a quadrupolar expansion of the {\it power spectrum} (PS) of the fluctuations \cite{Ackerman07}: $P_\zeta(\vec{k}) \equiv P_\zeta(k)\left[1+g_\zeta (k)(\hat{n} \cdot \hat{k} )^2 \right]$, where $g_{\zeta}(k)$ measures the magnitude of the statistical anisotropy and $\hat{n}$ points towards the preferred direction of the anisotropies \cite{Ackerman07,Groeneboom0,Groeneboom09,Kim13, Ramazanov:2013wea}.   We study the relation between the NG and the statistical anisotropy parameters and the sensitivity of the former to variations in the latter. We devote special attention to the scale dependence (or running) of the statistical anisotropy parameter $g_{\zeta}$  and the way in which it contributes to the shape and scale dependence of NG. Scale dependent NG \cite{Fergusson08, LoVerde07, Sefusatti09, Byrnes09, Byrnes10, Byrnes10b, Bartolo10a, Huang10, Huang11a, Huang11b, Tzavara10, Tzavara12, Byrnes12a} has been shown to be a generic feature of several inflationary models based on scalar fields such as Dirac-Born-Infeld (DBI) \cite{Chen05,Huang11b} and related models, so it is important to quantify this characteristic in order to have predictive power over the considered models \cite{Becker12}. Our main goal here is to quantify the relation between statistical anisotropy and NG and to identify how the {\it running of the statistical anisotropy level} is encoded in the {\it running and shape dependence of the NG parameters}.  Related studies about scale dependent anisotropic NG have been done in the context of noncommutative inflationary models without vector fields \cite{Nautiyal2013} and also in the context of LSS through scale dependent bias parameters induced by primordial vector fields \cite{Shiraishi13sv, Baghram13}. 
Throughout this paper, we employ the $\delta N$ formalism\footnote{Even though we use the $\delta N$ formalism, other formalisms can be used equivalently to obtain the same results.} \cite{Lyth:2005fi, Dimopoulos08} to obtain the statistics encoded in the correlation functions of the primordial curvature perturbation in inflationary models in the presence of vector fields. We will do a full analysis of the tree-level correlators and define a template for their quantitative evaluation extracting the scale and shape dependence on the wave vectors. 

This paper is organized as follows: In section \ref{psvf} we roughly review and discuss the general characteristics of inflationary models where vector fields play some role in the generation of the primordial curvature perturbation. In section \ref{correlators} we derive the correlation functions by means of the $\delta N$ formalism, identify the way in which the statistical anisotropy enters in the correlators and discuss some particular cases of interest.  The main results of this article are presented in section \ref{ng}; there, we evaluate the NG parameters in the presence of statistical anisotropy; we also discuss a way to track and identify the shape and scale dependence of the NG parameter $f_{\rm NL}$ by introducing the relevant definitions for measuring variations in the shape and size of the triangle formed by the wave vectors. Some numerical evaluations exemplifying and illustrating the way in which the statistical anisotropy affects the NG parameters are presented in section \ref{eval}. Finally,  in section \ref{results} we summarize our results and conclude.

\section{Primordial anisotropy in models which involve vector fields}\label{psvf}
The statistical anisotropy can be studied through the correlation functions of the primordial curvature perturbation $\zeta$. A common parametrization of the anisotropy in the power spectrum of $\zeta$ is obtained by doing a multipolar expansion, the lowest level being the quadrupolar term, as follows \cite{Ackerman07}:
\be\label{PSanis}
P_\zeta(\vec{k}) \equiv P_\zeta^{{\rm iso}}(k)\left[1+g(k)(\hat{n} \cdot \hat{k} )^2 \right],
\ee
where $\hat{n}$ is a unit vector pointing towards the preferred direction, $g$ is a function measuring the amplitude of the anisotropy, and $P_\zeta^{{\rm iso}}(k)$ denotes the isotropic part of the power spectrum. The five-year WMAP data were analyzed in order to obtain the value of $g$ and the orientation of the primordial anisotropy \cite{Groeneboom0,Groeneboom09}, and it was found that $g=0.290\pm 0.031$ nearly along the ecliptic poles. This result was confronted later in Refs. \cite{Hanson09, Hanson10}, and it was claimed that this value might have its origin in systematic errors in the measurements due to asymmetric beams in the instruments. The PLANCK mission at first seemed to confirm the presence of anomalies due to statistical anisotropy \cite{PlanckSI13}, but soon after, using the PLANCK 2013 temperature map results, it was shown that there is ``{\it no evidence for violation of rotational symmetry}''  with $g= 0.002 \pm 0.016$ (68 $\%$ {\rm CL}) \cite{Kim13}, where the asymmetry beam effects were removed;  the latter result is consistent with the analysis performed in Ref. \cite{Ramazanov:2013wea} employing the WMAP nine-year data.

Despite these results disfavoring the presence of significant statistical anisotropy in the power spectrum, it is still possible that it appears in higher order correlation functions. To this end, in this paper we also study how the statistical anisotropy  appears in the three-point correlator, parametrized by the bispectrum (BS), and its effects on the non-Gaussianity parameter   $f_{\rm NL}$, especially on its scale and shape dependence. The analysis of the non-Gaussianity parameters $g_{{\rm NL}}$ and $\tau_{{\rm NL}}$ related to the four-point correlator, parametrized by the trispectrum, and higher order parameters is intricate but straightforward.

Several models produce statistical anisotropy which is parametrized as in Eq. (\ref{PSanis}). We consider models which can be derived from the action
\be\label{ssv}
S= \int d^{4}x \ \sqrt{-g} \ {\cal{L}}(R, \phi, A_\mu) \,,
\ee 
with a single scalar inflaton field $\phi$ and a vector field $A_i$ both minimally coupled to Einstein gravity. In principle, we can consider a vector field with a potential term (or simply, a mass term) but, at some point, we shall specialize in the massless $\rm{U}(1)$ invariant case which is ghost instability free \cite{Carroll:2009em}; an example of a massive, stable model, however, has been presented in Refs. \cite{Dimopoulos09a,Dimopoulos09vu}.  Several models of this type have been studied with interest in recent years; among them, it is worth mentioning the model with the coupling term $f^{2}(\phi)F^{\mu \nu}F_{\mu \nu}$  \cite{Watanabe:2009ct}
mainly because it is a stable model able to produce both statistical anisotropy and anisotropic expansion which do not dilute during the inflationary expansion (see, for instance, Refs. \cite{Maleknejad:2011jr,Bartolo12,Lyth13} and references therein).  
 
Roughly speaking, one can solve the equations of motion derived from Eq. (\ref{ssv}) for the scalar field perturbations $\delta \phi$ and vector field perturbations $\delta A_{i}$  and express the general solution as an expansion: 
\ba
\delta \phi &=& \sum_{\vec{k}} \left[ \hat{a}(\vec{k}) f(\tau, k) +  \hat{a}^{\dagger}(-\vec{k}) f^{*}(\tau, k) \right] e^{i\vec{k}\cdot \vec{x}} \,, \\
\delta A_{j} &=& \sum_{\vec{k} , \lambda} \left[ {\bf e}_{j\lambda}(\hat{k}) \hat{a}_{ \lambda}(\vec{k}) w_{ \lambda}(\tau, k)  +   {\bf e}^{*}_{j\lambda}(-\hat{k}) \hat{a}^{\dagger}_{ \lambda}(-\vec{k}) w^{*}_{ \lambda}(\tau, k) \right] e^{i\vec{k}\cdot \vec{x}} \,, 
\ea
where $\tau$ is the cosmic time, the functions $f$ and $w_{ \lambda}$ form, respectively, a complete basis for the solutions of the equations of motion, the vectors ${\bf e}_{i\lambda}$ are the polarization vectors of vector perturbations, and $\hat{a}(\vec{k}), \hat{a}^{\dagger}(\vec{k})$ and  $\hat{a}_{\lambda}(\vec{k}), \hat{a}^{\dagger}_{\lambda}(\vec{k})$ are the creation and annihilation operators for scalar and vector field perturbations, respectively, such that they satisfy $ \hat{a}(\vec{k})|0 \rangle = 0,   \hat{a}_{\lambda}(\vec{k})|0 \rangle =0$, and obey the commutation rules:
\be
\left[ \hat{a}(\vec{k}), \hat{a}^{\dagger}(\vec{k'}) \right] =  \delta(\vec{k}-\vec{k}') \,, \;\;\; \left[ \hat{a}_{\lambda}(\vec{k}), \hat{a}^{\dagger}_{\lambda'}(\vec{k'}) \right] =  \delta_{\lambda \lambda'} \delta(\vec{k}-\vec{k}') \,.
\ee
The polarizations of the vector field perturbations are the transverse right (R) and left (L) and the longitudinal (long) in the massive case, so $\lambda= \{ {\rm R}, {\rm L}, {\rm long} \}.  $
By choosing standard vacuum  conditions and using the several asymptotic properties of Hankel and Bessel functions, one can express the solutions up to first order in slow-roll parameters as follows\footnote{This form is a direct generalization of the results for multiscalar field inflation shown in Refs. \cite{Nakamura96, Byrnes09, Byrnes10} and the results in Ref. \cite{Dimopoulos09a} for a vector curvaton field with a varying kinetic function during inflation.}:
\be\label{scale_pert}
\delta \Phi_{I} (\vec{k})= \frac{i H}{\sqrt{2k^{3}}}\left[ (1-\epsilon)\delta^{J}_{I} + \left[ c+\ln\left(\frac{aH}{k}\right)  \right] \epsilon^{J}_{I} \right] \sum_{\lambda} {\bf e}_{J \lambda}(\hat{k})\hat{a}_{\lambda}(\vec{k}) \,.
\ee
In the previous expression, we have introduced the compact notation  $\delta \Phi_{I} = (\delta \phi\, ,\delta A_{i})$, where the index $I$ labels both the scalar field and the components of the vector field.  In addition, $H$ is the Hubble parameter during inflation, $a$ is the expansion parameter, 
\ba 
\epsilon &\equiv& - \frac{\dot{H}}{H^{2}} \,, \\  
\epsilon_{IJ} &\equiv& \epsilon \delta_{IJ} + \Sigma_{IJ} \,, 
\ea
$c= 2-\ln 2 - \gamma = 0.7296\dots$, with $\gamma$ being the Euler-Mascheroni constant, and the polarization vectors also follow a compact notation\footnote{In this compact notation, $a_0(\vec{k})$ means $a(\vec{k})$, i.e., the annihilation operator for the scalar perturbation.}: ${\bf e}_{\lambda}^{J}=\{ \delta_{\lambda}^{0}, {\bf e}_{\lambda}^{i}\}$. Regarding the $\Sigma_{IJ}$ symbols, they are related to the slow-roll parameters, their precise form depending on the particular model; for instance, in multiscalar field models one gets 
\be
\Sigma_{IJ} =  \frac{V_{I} V_{J}}{9H^{4}} -   \frac{V_{IJ} }{3H^{2}} \,,
\ee
where $V_{I} = \frac{\partial{V}}{\partial{\phi_{I}}}$ and $V_{IJ} = \frac{\partial^{2}{V}}{\partial{\phi_{I}} \partial{\phi_{J}}}$.  
\section{Scale and shape dependent statistical anisotropy in the correlation functions}\label{correlators}
In cosmological inflationary models that include vector fields, part, or even all, of the curvature perturbation can be generated by the vector field perturbations.  In such scenarios, $\zeta$ can be calculated through the $\delta N$ formalism \cite{Dimopoulos08,Lyth:2005fi}  which states that one can evaluate the primordial curvature perturbation $\zeta$ in terms of the derivatives of the amount of expansion  $N$ with respect to the background field values, and the values of the field perturbations, in the flat slicing, at the time of horizon exit\footnote{Since the gauge is fixed so that $A_0 = 0$ (in the background), the perturbation $\delta A_0$ does not enter in Eq. (\ref{deltaN}).}. The expression that we use for our calculations is the following \cite{Valenzuela11}:  
\ba  \label{deltaN}
\zeta (\vec{x}, t) \equiv \delta N(\vec{x}, t) 
= N_{I} \delta \Phi_I  + \frac{1}{2!}N_{ I J } \delta \Phi_I \delta \Phi_J +\frac{1}{3!}N_{ I J K } \delta \Phi_I \delta \Phi_J \delta \Phi_K + \cdots \,.
\ea 

For the amount of expansion derivatives we use
\be 
N_{I} \equiv \frac{\partial N}{\partial \Phi_{I}}, \; N_{IJ} \equiv \frac{\partial^{2} N}{\partial \Phi_{I}\partial \Phi_{J}}, \; \mbox{etc} \,.
\ee
The $N$ derivatives are evaluated at the initial time $t$ at some instant soon after horizon exit defined by the scale $k = a(t^*) H(t^*)$ ($t^*$ being the time of horizon exit), so they carry some scale dependence also encoded in this time dependence. The scale dependence in the derivatives can be calculated, for instance, as explained in Ref. \cite{Sasaki95} by evaluating the derivative 
\be \frac{\partial }{\partial \ln k}\Big|_{a = \rm constant} = \frac{\partial }{\partial \ln a}\Big|_{k/a = \rm constant} - \frac{\partial }{\partial \ln a}\Big|_{k = \rm constant} \,,\ee
or as derived at second order in slow-roll parameters in Ref. \cite{Nakamura96}.

With the curvature perturbation expansion in Eq. (\ref{deltaN}), we can calculate the different correlation functions and see how the scale and shape dependence enters in the statistical parameters. 
\subsection{Power spectrum}
We start with the PS:
\be\label{2p}
\langle \zeta({\vec k}_1) \zeta({\vec k}_2) \rangle \equiv (2\pi)^3 \delta({\vec k}_{12}) P_{\zeta}({\vec k}_1)
\equiv (2\pi)^3 \delta({\vec k}_{12}) \frac{2 \pi^2 }{k^3}{\cal P}_{\zeta} ({\vec k}_1) \,,
\ee
where ${\cal P}_{\zeta} ({\vec k})$ is the {\it dimensionless power spectrum}, and ${\vec k}_{12} \equiv {\vec k}_{1} + {\vec k}_{2}$. 
We shall restrict our analysis  by doing some assumptions about the field perturbations. First, we consider that the expansion is Friedmann-Robertson-Walker (FRW) type.  Second, NG in the field perturbations is negligible; that is, the NG is generated due to the superhorizon evolution.  This will be important since, this being the case, the primordial NG can be big enough to be observable by foreseeable dedicated missions and, as we will see, the corrections due to the statistical anisotropy can, in principle, be too. Third, we assume that the correlations between scalar and vector field perturbations at horizon exit are subleading  order \cite{Valenzuela11}. Fourth, we consider that the vector field perturbations do not violate parity symmetry. We shall adopt the notation
\be
 \langle \delta \Phi_I (\vec {k}_1)  \delta \Phi_J(\vec {k}_2)  \rangle \equiv (2\pi)^3 \delta({\vec k}_{12}) \Pi_{IJ}(\vec{k}_1) \,,
\ee
with the components
\ba \langle \delta \phi (\vec {k}_1)  \delta \phi (\vec {k}_2)  \rangle &\equiv& (2\pi)^3 \delta({\vec k}_{12})  P_{\delta \phi}({k}_1) \,, \\ 
\langle \delta A_i (\vec {k}_1)  \delta A_j(\vec {k}_2)  \rangle &\equiv& (2\pi)^3 \delta({\vec k}_{12}) \Pi_{ij}(\vec{k}_1) \equiv (2\pi)^3 \delta({\vec k}_{12})\sum_{\lambda}{\bf e}_{i}^{\lambda}(\hat{k}_{1}){\bf e}_{j}^{\lambda *}(-\hat{k}_{1})P_{\lambda}(\vec{k}_{1}) \,,
\ea
and zero otherwise. In the latter expression,
\begin{equation}
\delta A_i (\vec {k}) \equiv \sum_{\lambda= {\rm R, L, long}}{\bf e}_{i}^{\lambda}(\hat{k})\delta A_{\lambda} (\vec {k}) \,,
\end{equation}
and
\begin{equation}
\sum_{\lambda}{\bf e}_{i}^{\lambda}(\hat{k}){\bf e}_{j}^{\lambda *}(-\hat{k})P_{\lambda}(\vec{k}) \equiv \Pi_{ij}(\vec{k}) = \Pi^{\rm even}_{ij}(\vec{k}) P_{+}(k) + \Pi^{\rm odd}_{ij}(\vec{k}) P_{-}(k) + \Pi^{\rm long}_{ij}(\vec{k}) P_{{\rm long}}(k) \,,
\end{equation}
where $\Pi^{\rm even}_{ij}(\vec{k}) \equiv \delta_{ij} - \hat{k}_{i}\hat{k}_{j}, \ \Pi^{\rm odd}_{ij}(\vec{k}) \equiv i \epsilon_{ijk}\hat{k}_{k}$, and
$\Pi^{\rm long}_{ij}(\vec{k}) \equiv  \hat{k}_{i}\hat{k}_{j}$ \cite{Dimopoulos08,Valenzuela11}.  $P_{\rm long}(k)$ is the longitudinal (``massive component") part of the power spectrum and $P_{+}(k)$ and $P_{-}(k)$ are, respectively, the parity conserving and parity violating power spectra: $P_{{\pm}}=(P_{R}\pm P_{L})/2$ \cite{Valenzuela11,Dimopoulos08}. The scale dependence in every spectra can be derived from the analysis sketched in section \ref{psvf} that leads to Eq. (\ref{scale_pert}), and their precise forms depend on the specific details of the model. We do not enter in the specifics of each model but, instead, we consider that the scale dependence in every spectrum can be written as a power law, which is quite a general consideration. 

Additionally, given the separate universe assumption \cite{Lyth:2009zz}, which the $\delta N$ formalism is based on, the form of the equations for the dynamical quantities at each comoving location is the same as for the unperturbed quantities.  Thus, the field perturbation equations in momentum space do not depend explicitly on $\vec{k}$.  They depend on time, thus on $k_*$ (the wavenumber at horizon exit), but they do not depend on the direction of $\vec{k}$.  Their solutions are, therefore, independent of the direction of $\vec{k}$ except for the set of initial conditions $\{\alpha^n_{\vec{k}}\}$.  However, the field perturbations are evaluated in the flat slicing, so, taking into account the assumption of a FRW-type expansion, the whole perturbed metric in this slicing is actually FRW which is conformally equivalent to Minkowski.  As a consequence, the set of initial conditions, when quantizing, can be written as $\{\hat{\alpha}^n_{\vec{k}} = \alpha^n_k \hat{a}^n_{\vec{k}}\}$, where $\hat{a}^n_{\vec{k}}$ is the respective annihilation operator.  Following the usual procedure to calculate the power spectrum of the field perturbations, this implies that none of $P_{\delta \phi}(k)$, $P_{+}(k)$, $P_{-}(k)$, and $P_{\rm long}(k)$ depends on the direction of the wave vector (since the $\alpha^n_k$ do not depend on the direction of $\vec{k}$).  This is valid for all the relevant cosmological scales since the time of horizon exit for each of them is contained within the time interval spanning from the beginning of inflation to the time when the curvature perturbation $\zeta$ is evaluated.  Thus, at tree level, the PS is \cite{Valenzuela11,Dimopoulos08}
\be \label{ps-zeta}
P_{\zeta}(\vec{k}) = N_{I} N_{J} \Pi_{IJ}(\vec{k}) = (N_{\phi})^{2}  P_{\delta\phi}({k}) +  N_{i} N_{j} \Pi_{ij}(\vec{k})  \,.
\ee


As already mentioned,  in this article we only consider the case of parity conserving vector field perturbations; then, we only  keep $P_{+}(k)$ and  $P_{{\rm long}}(k)$. We need not suppose that all the $\delta A_{i}$ perturbations evolve in the same way after horizon exit so that they do not necessarily have the same scale dependence on their spectra. With these assumptions, the longitudinal and the parity conserving  spectra can be related by $P_{{\rm long}}(k) = q(k) P_{+}(k)$ where the $q(k)$ function could, in general situations, carry some level of scale dependence. Then, the vector power spectra read
\be \label{vps}
\Pi_{ij}(\vec{k}) = \left[ \delta_{ij} + \left(q(k)-1\right) \hat{k}_{i}\hat{k}_{j}\right] P_{+}(k) \,.
\ee
Therefore, the PS obtained from the $\delta N$ formalism is
\ba
\label{ps-zeta-tilde}
{ P}_{\zeta}(\vec{k})= { P}^{\rm iso}_{\zeta}(k) + { P}^{\rm aniso}_{\zeta} (\vec{k} )  = { P}^{\rm iso}_{\zeta}(k)\left[1  +  g_\zeta (k)(\hat{n}_{i} \hat{k}_{i})^2 \right] \,,
\ea
which fits in the parametrization given in Eq. (\ref{PSanis}). In the previous expression we must employ 
\begin{eqnarray}
{ P}^{\rm iso}_{\zeta} (k) &\equiv&  \frac{2\pi^{2}}{k^{3}}  {\cal P}^{\rm iso}_{\zeta}(k) = {N}_{I} {N}_{J} { P}_{IJ}(k)  \,, \label{isops} \\
{ P}^{\rm aniso}_{\zeta} (\vec{k}) &=& (q(k)-1)\left({N}_{i} \hat{k}_{i} \right)^2 { P}_{+}(k) \,, \label{anisops}
\end{eqnarray}
and
\be
\hat{n}_i \equiv N_i/(N_j N_j)^{1/2} \,.
\ee
The relation between the $(q(k)-1)$ factor and the anisotropy parameter $g_\zeta(k)$ in the power spectrum is given by
\be\label{g-anis}
g_\zeta(k)= (q(k)-1)\frac{{N_i N_i} { P}_{+} (k) }{{N}_{I}  {N}_{J} { P}_{IJ}(k)} = (q(k)-1)\frac{ {\cal P}_{\zeta_{A_{+}}}(k) }{ {\cal P}^{\rm iso}_{\zeta}(k) } \,,
\ee
where the power spectra $ { P}_{IJ}$ assume the values $ { P}_{00} =  { P}_{\delta \phi}$ and  ${ P}_{ij}  =  \delta_{ij}{ P}_{+}$, and the contributions to the spectrum of $\zeta$ from the scalar and the vector field are defined as follows:
\begin{eqnarray}
P_{\zeta_{\phi}} (k) &\equiv& \frac{2\pi^{2}}{k^{3}}  {\cal P}_{\zeta_{\phi}} (k) = N_{\phi}^{2} { P}_{\delta \phi}(k) \,, \\
P_{\zeta_{A_{+}}} (k) &\equiv&  \frac{2\pi^{2}}{k^{3}}  {\cal P}_{\zeta_{A_{+}}} (k) = N_{i} N_{i} { P}_{+}(k) \,. 
\end{eqnarray}

We must notice that, according to Eq. (\ref{g-anis}), generically, the simultaneous presence of scalar and vector perturbations induces scale dependence in $\gz$.  At this point, we assume that the isotropic part of the spectrum of the primordial curvature perturbation can be expressed as a power law where deviations from scale invariance are parametrized by a spectral index $n^{{\rm iso}}_{\zeta}$. The precise forms of the spectral index and its running, as we said before, depend on the slow-roll parameters in the model but, for our purposes, we do not need a concrete expression. Then, the isotropic spectrum (which includes both scalar and vector perturbations) is written as
\be {\cal P}^{\rm iso}_{\zeta}(k) =  {N}_{I}  {N}_{J} {\cal P}_{IJ}(k) \equiv {\cal A_{\zeta}}  \left(\frac{k}{k^{*}}\right)^{n^{{\rm iso}}_{\zeta}-1} \,, \ee
with an amplitude ${\cal A_{\zeta}}$ at the pivot scale $k^{*}$. In the same way, the spectrum of the scalar perturbations is parametrized as 
\be   {\cal P}_{\zeta_{\phi}}(k) =  {N_\phi^{2}} {\cal P}_{\phi} (k) \equiv {\cal A_{\phi}}  \left(\frac{k}{k^{*}}\right)^{n_{\phi}-1} \,, \ee
and the spectra of the longitudinal and transverse polarizations of the vector perturbations become
\begin{eqnarray}   
{\cal P}_{\zeta_{A_{+}}}(k) &=&  {N_i N_i} {\cal P}_{+} (k) \equiv {\cal A_{+}}  \left(\frac{k}{k^{*}}\right)^{n_{+}-1} \,, \\
{\cal P}_{\zeta_{  A_{\rm{ long}}}}(k) &\equiv&  {\cal A_{\rm{long}}}  \left(\frac{k}{k^{*}}\right)^{n_{\rm{long}}-1} \,,  
\end{eqnarray}
 where the spectral indices $n_{+}$  and $n_{\rm long}$ can run with the scale. If we evaluate the logarithmic derivative of the power spectrum in Eq. (\ref{ps-zeta-tilde}), we can define a spectral index for this class of models:
 \ba \label{nps}
 n_{\zeta} -1 &\equiv&  \frac{d \ln {\cal P}_{\zeta}(\vec{k}) }{d \ln k } = \frac{d \ln {\cal P}^{\rm iso}_{\zeta}({k}) }{d \ln k } - \frac{1}{1 +  g_\zeta (k)(\hat{n}_{i} \hat{k}_{i})^2 } \frac{d \ln g_{\zeta}({k}) }{d \ln k } \\
                  &\equiv&  n^{\rm iso}_{\zeta} - 1 - \frac{1}{1 +  g_\zeta (k)(\hat{n}_{i} \hat{k}_{i})^2 } ( n_g - 1) \,,
 \ea
where, according to the definitions given above,
\be
 \frac{d \ln {\cal P}^{\rm iso}_{\zeta}({k}) }{d \ln k } = n^{\rm iso}_{\zeta} - 1 = \frac{1}{ {\cal P}_{\zeta}^{\rm iso} } \left[ (n_{+} -1){\cal P}_{\zeta_{A_{+}}} + (n_{\phi} -1){\cal P}_{\zeta_{\phi}} \right] \,.
\ee 
The quantity  $ n_g - 1 \equiv \frac{d \ln g_{\zeta}({k}) }{d \ln k }$ measures the deviation from scale invariance of the anisotropy parameter function $g_{\zeta}(k)$. We see from the above expressions that the spectral index $n_\zeta$ is sensitive to the preferred orientation defined by the vector $\hat{n}$ and, certainly, the latter plays a role for quantifying and measuring deviations from scale invariance and statistical isotropy. Now, we can think about some simplified scenarios; for instance, we can consider that the longitudinal and transverse spectral indices of the vector perturbations are equal, so that the $q$ ratio is a constant. With this in mind, it is easy to realize that the statistical anisotropy parameter $g_{\zeta}$ can be expressed as a power law: 
\be\label{gpl}
g_\zeta(k)\equiv   g_{\zeta}^{*} h(k) = g_{\zeta}^{*} \left(\frac{k}{k^{*}}\right)^{n_{g} - 1} \,, 
\ee
where
\be g_{\zeta}^{*}  = (q-1) \frac{{\cal A_{+}}}{ {\cal A_{\zeta}}} \,,
\ee
and
\be
n_{g}  - 1 = n_{+} - n^{\rm iso}_{\zeta} \,.\ee

Later, we will devote special attention to an important particular case: the massless vector field.  In that case, the longitudinal polarization is null;  then, the $q$ ratio is zero and the statistical anisotropy parameter is
\be g_{\zeta}(k) = - \frac{ {\cal P}_{\zeta_{A_{+}}}(k) }{ {\cal P}^{\rm iso}_{\zeta}(k) } = - \frac{{\cal A_{+}}}{ {\cal A_{\zeta}}}   \left(\frac{k}{k^{*}}\right)^{n_{+} - n^{\rm iso}_{\zeta}} \,. \ee 
A remarkable fact is that, in the massless case, the statistical anisotropy factor is always negative and its running can be deduced directly from the running of the scalar and the transverse vector perturbation spectra. It is also worth noticing that, in this case, $g_{\zeta}  \geqslant -1$.

\subsection{Bispectrum}
Now, we follow the same procedure for the evaluation of the BS:
\be
\langle \prod_{i=1}^3\zeta({\vec k}_i)\rangle = (2\pi)^3 \delta ({\vec k}_{123}) B_{\zeta}({\vec k}_1 ,\,{\vec k}_2 ,\, {\vec k}_3 ), 
\ee
where ${\vec k}_{123} = {\vec k}_1 + {\vec k}_2 + {\vec k}_3$.

We get, from the $\delta N$ formula, the tree-level expression
\ba 
B_{\zeta}(\vec{k}_{1}, \vec{k}_{2}, \vec{k}_{3})= N_{I} N_{J} N_{K} B_{IJK}(\vec{k}_{1}, \vec{k}_{2}, \vec{k}_{3})  + N_{I} N_{J} N_{KL}\left[ \Pi_{IK}(\vec{k}_1) \Pi_{JL}(\vec{k}_2) + {\mbox  {\rm cyc. perm.}} \right] \,, \nonumber \\
\label{bs-zeta}
\ea
where $B_{IJK}(\vec{k}_{1}, \vec{k}_{2}, \vec{k}_{3})$ is the connected part of the three-point correlator and its  components describe the primordial BS for the different field perturbations at the time of horizon exit. As we did with the PS, now we will impose some extra simplifying conditions and discuss their validity under some general grounds. First, we suppose that the derivatives of $N$ obey
\be\label{ninij} 
N_{I} \propto \Phi_{I}, \; N_{IJ}  \propto \delta_{IJ}. 
\ee
The second condition involving the second order derivatives implies that the mixed interactions between scalar and vector fields  ($N_{i \phi}$)  are subleading order compared to the scalar-scalar and vector-vector interactions.
Although it is not essential, we invoke this condition just to simplify the calculations and because those mixed terms do not  seem to add significant contributions to the correlations (see, however, Ref. \cite{Namba12gg} for a detailed calculation of the PS in the $f^2(\phi) F^{\mu \nu}F_{\mu \nu} +m^2(\phi)A^2 $ curvaton model considering  the mixed scalar and vector terms). Some of the most studied inflationary models in the presence of vector fields, such as the vector curvaton \cite{Dimopoulos06}, vector inflation \cite{Golovnev08}, hybrid inflation \cite{Yokoyama08} and their variants, obey the conditions in Eq. (\ref{ninij}) to a good approximation and can be parametrized as we do here.

With these considerations, the bispectrum reads
\ba\label{bs0}
 B_{\zeta} & \equiv &  (1+\xi_1)B_{\zeta}^{{\rm iso}} \\ \nonumber 
& = & \,\left[ 1 +   \,\frac{ {\sum\limits_{l<m}\, 
N_{+}^{2}N_{++} P_{+}(k_{l}) P_{+}(k_{m}) \left[{ Q(k_{l})  }(\hat n\cdot \hat k_{l} )^{2}  + Q(k_{m})(\hat n\cdot \hat k_{m} )^{2} \right]} }{ N_{A}N_{B} N_{CD} \sum\limits_{l<m} P_{AC}(k_{l}) P_{BD} (k_{m}) } \, \right. \\ \nonumber 
 &+&   \left.  \frac{ {  \sum\limits_{l<m}
N_{+}^{2}N_{++} P_{+}(k_{l}) P_{+}(k_{m}) Q(k_{l}) Q(k_{m}) (\hat n\cdot \hat k_{l}) (\hat n\cdot \hat k_{m}) (\hat k_{l} \cdot \hat k_{m})  } }{ N_{A}N_{B} N_{CD} \sum\limits_{l<m} P_{AC}(k_{l}) P_{BD} (k_{m}) }\right] B_{\zeta}^{{\rm iso}} \,,
\ea 
where we have introduced the definitions
\be
Q(k) \equiv q(k) -1 \,, \qquad N_{i} \equiv  N_{+} {\hat n}_{i} \,, \qquad   N_{ij} \equiv  N_{++} \delta_{ij} \,,
\ee
and
\ba\label{bs_vec+scalar}
B_{\zeta}^{{\rm iso}}({ k}_1 ,\,{ k}_2 ,\, { k}_3 ) &=&  N_{I} N_{J} N_{KL}\left[ P_{IK}({k}_1) P_{JL}({k}_2) + {\mbox {\rm cyc. perm.}} \right] \\ \nonumber
&=& \frac{N_{++}}{N_{+}^{2}} P_{\zeta_{A_{+}}}(k_{1}) P_{\zeta_{A_{+}}}(k_{2}) + \frac{N_{\phi \phi}}{N_{\phi}^{2}} P_{\zeta_{A_{\phi}}}(k_{1}) P_{\zeta_{A_{\phi}}}(k_{2})  + {\mbox{\rm cyc. perm.}} 
\ea
In the previous expressions, the scale dependence coming from the statistical anisotropy parameter $g_\zeta(k)$ enters through the longitudinal to transverse factor $q$ in the function $Q(k) = q(k) -1$. We stress that the only assumption that we have made in deriving Eq. (\ref{bs0}) is that, under the considerations and approximations that we discussed before, the second derivatives of the amount of expansion $N$ are such that the mixed scalar-vector components are subleading order compared to the scalar-scalar and the vector-vector components; that is, $N_{i\phi}\approx 0$.  

Now, one can use Eq. (\ref{g-anis}) to rewrite the BS also in terms of the $g_{\zeta}$ parameter instead of the $q$ ratio to have a direct relation between the BS and the statistical anisotropy parameter in the PS. This will allow us to identify the functional dependence of the non-Gaussianity parameter $f_{\rm NL}$ in terms of the statistical anisotropy level $g_{\zeta}$ and, therefore, to isolate the scale dependence encoded only on the $g_\zeta$ parameter. The result is
\ba\label{bs-g}
 B_{\zeta} & \equiv &  (1+\xi_1)B_{\zeta}^{{\rm iso}}  \\ \nonumber 
&=& \,\left[ 1+ \,  \,\frac{ {\sum\limits_{l<m}\, 
 \left( \frac{N_{++} }{ N_{+}^{2} } \right) \left[   P_{\zeta_{A_{+}}}(k_{m})  P^{\rm iso}_{\zeta} (k_{l})    g_{\zeta}(k_{l}) (\hat n\cdot \hat k_{l} )^{2}  +    P_{\zeta_{A_{+}}}(k_{l})  P^{\rm iso}_{\zeta} (k_{m})  g_{\zeta}(k_{m})   (\hat n\cdot \hat k_{m} )^{2} \right]} } { N_{A}N_{B} N_{CD} \sum\limits_{l<m} P_{AC}(k_{l}) P_{BD} (k_{m}) }\, \right. \\ \nonumber 
 &+&   \left.  { \frac{ \sum\limits_{l<m} 
 \left( \frac{N_{++} }{ N_{+}^{2} } \right) P^{\rm iso}_{\zeta}(k_{l}) P^{\rm iso}_{\zeta}(k_{m}) g_{\zeta}(k_{l}) g_{\zeta}(k_{m}) (\hat n\cdot \hat k_{l}) (\hat n\cdot \hat k_{m}) (\hat k_{l} \cdot \hat k_{m})  }{ N_{A}N_{B} N_{CD} \sum\limits_{l<m} P_{AC}(k_{l}) P_{BD} (k_{m})}} \right] B_{\zeta}^{{\rm iso}} \,,
\ea 
where the $1 + \xi_1$ parameter has been defined and corresponds to the expression inside the big brackets. 
We can also extend our analysis to higher order NG parameters, but our main focus here is the three-point correlator. Higher order cases are straightforward to generalize, but their calculations are rather lengthy and intricate and do not contribute significantly to the analysis presented here.  

\section{Scale and shape dependent non-Gaussianity and statistical anisotropy }\label{ng}
A legitimate question, which one would like to answer when considering models with statistical anisotropies, is if there is any relation between the NG parameters and the level of statistical anisotropy in the statistical distribution of the fluctuations. In this section, we answer affirmatively to this question and describe quantitatively the relation between NG and statistical anisotropy, considering also the scale (running) and shape dependence of the correlators in the presence of inflationary vector fields.   
\subsection{Scale and shape dependence in $f_{\rm NL}$}
The $f_{\rm NL}$  NG parameter is defined by
\be\label{fnl}
\frac{6}{5}f_{\rm NL}({\vec k_1}, {\vec k_2}, {\vec k_3}) \equiv \frac{B_{\zeta}({\vec k_1}, {\vec k_2}, {\vec k_3})}{P_{\zeta}({\vec k_1})P_{\zeta}({\vec k_2})+{\mbox {\rm cyc. perm.}}} \,.
\ee
To calculate it, we also need the cyclic permutations of the PS products in the denominator. Using Eqs. (\ref{ps-zeta-tilde}) and (\ref{gpl}) we obtain
\ba\label{psps} 
&& P_{\zeta}({\vec k_1})P_{\zeta}({\vec k_2})+{\mbox {\rm cyc. perm.}} \equiv (1+\chi_1) \left[P^{{\rm iso}}_{\zeta}({ k_1})P^{{\rm iso}}_{\zeta}({ k_2})+{\mbox {\rm cyc. perm.}}\right]  \\  \nonumber 
&& \equiv \,\left[ 1+ \,\frac{ N_{I} N_{J} N_{K} N_{L} \sum\limits_{l<m}\, 
P_{IJ}({ k_l})P_{KL}({ k_m})  \left[ g_{\zeta}(k_{l})(\hat n\cdot \hat k_{l} )^{2}  + g_{\zeta}(k_{m})(\hat n\cdot \hat k_{m} )^{2} \right]}{  N_{A} N_{B} N_{C} N_{D} \sum\limits_{l<m}
P_{AB}({ k_l})P_{CD}({ k_m})  } \, \right. \\  \nonumber
&& +   \left. \,  \frac{ N_{I} N_{J} N_{K} N_{L} \sum\limits_{l<m}\, 
P_{IJ}({ k_l})P_{KL}({ k_m})  g_{\zeta}(k_{l}) g_{\zeta}(k_{m}) (\hat n\cdot \hat k_{l} )^{2} (\hat n\cdot \hat k_{m} )^{2} }{   N_{A} N_{B} N_{C} N_{D} \sum\limits_{l<m}
P_{AB}({ k_l})P_{CD}({ k_m})  } \right]  \\ \nonumber
&& \times N_{A} N_{B} N_{C} N_{D} \sum\limits_{l<m}
P_{AB}({ k_l})P_{CD}({ k_m})  \\ \nonumber 
&& = \,\left[ 1+ \,\frac{\sum\limits_{l<m}\, 
P^{{\rm iso}}_{\zeta}({ k_l})P^{{\rm iso}}_{\zeta}({ k_m})  \left[ g_{\zeta}(k_{l})(\hat n\cdot \hat k_{l} )^{2}  + g_{\zeta}(k_{m})(\hat n\cdot \hat k_{m} )^{2} \right]}{   \sum\limits_{l<m}
P^{{\rm iso}}_{\zeta}({ k_l})P^{{\rm iso}}_{\zeta}({ k_m})  } \, \right. \\  \nonumber
&& +   \left. \,  \frac{\sum\limits_{l<m}
P^{{\rm iso}}_{\zeta}({ k_l})P^{{\rm iso}}_{\zeta}({ k_m})  g_{\zeta}(k_{l}) g_{\zeta}(k_{m}) (\hat n\cdot \hat k_{l} )^{2} (\hat n\cdot \hat k_{m} )^{2} }{   \sum\limits_{l<m}
P^{{\rm iso}}_{\zeta}({ k_l})P^{{\rm iso}}_{\zeta}({ k_m}) } \right] \sum\limits_{l<m}
P^{{\rm iso}}_{\zeta}({ k_l})P^{{\rm iso}}_{\zeta}({ k_m}) \,,
\ea
where the $1 + \chi_1$ parameter has been defined and it corresponds to the expression inside the big brackets.
Now, with the expressions in Eqs. (\ref{bs-g}) and (\ref{psps}) we can go into the details of the computation of the  $f_{\rm NL}$ parameter for some inflationary models including vector fields. We track the scale dependence of the anisotropy parameter $g_{\zeta}(k)$ and its effect on the scale and shape dependence of $f_{\rm NL}$.
The class of models that we consider in this section can be parametrized using its first three correlation functions. We recall that, for simplicity, we shall consider models in which there is only one vector field, so there is only one preferred direction given by the unitary vector $\hat{n}_{i} = N_{i}/(N_{k}N_{k})^{1/2}$, the generalization for several scalar and vector fields being straightforward (for more details, see Ref. \cite{Valenzuela11}). 
Having this in mind, the $f_{\rm NL}$ parameter can then be written as
\be\label{ftofiso}
\frac{6}{5}f_{\rm NL}({\vec k_1}, {\vec k_2}, {\vec k_3}) = \frac{(1+\xi_1)}{ (1+\chi_1)}  \frac{  B^{\rm iso}_{\zeta}(k_{1}, k_{2}, k_{3}) }{ \sum\limits_{l<m}
P^{{\rm iso}}_{\zeta}({ k_l})P^{{\rm iso}}_{\zeta}({ k_m}) }= \frac{(1+\xi_1)}{ (1+\chi_1)}\frac{6}{5}f^{\rm iso}_{\rm NL}({ k_1}, { k_2}, { k_3}) \,,
\ee
where $f^{\rm iso}_{\rm NL}$ corresponds to the isotropic part of the full $f_{\rm NL}$ parameter. In general, $f_{\rm NL}$ will carry some scale dependence due to the length of each side of the momenta triangle, i.e., $k_1,\ k_2,$ and $k_3;$  when allowing for statistical anisotropy, it also carries extra dependence due to both the orientation of the referred triangle in momentum space, defined by $\hat{k}_1,\ \hat{k}_2,$ and $\hat{k}_3,$ and the orientation of the vector $\hat{n}$.  Although all of these parameters, $k_1,\ k_2,\ k_3,\ \hat{k}_1, \ \hat{k}_2,\ \hat{k}_3,	$ and $\hat{n}$,  appear both in $\zeta_1$ and $\chi_1,$ these two parameters actually depend only on $\vec{k}_1,\ \vec{k}_2,\ \vec{k}_3,$ and $\hat{n}$ (since they parametrize deviations from statistical isotropy);  meanwhile, $f^{\rm iso}_{\rm NL}$ only receives contributions from $k_1,\ k_2,$ and $k_3$.
 In order to deal with the scale and shape dependence, we are going to employ the useful and convenient set of variables introduced in Refs. \cite{Rigopoulos04, Fergusson08}. They will allow us to identify and separate genuine scale effects related to the size of the momenta triangle from nongenuine scale effects related to the shape of the same\footnote{By nongenuine scale effects we mean that, although $\alpha_1$ and $\alpha_2$ in Eq. (\ref{pshape}) seem to depend explicitly on $k_1,\ k_2,$ and $k_3$ (i.e., on the size of the triangle), they actually do not since the shape of the triangle is preserved under scaling of its size.}.  The ``size'' is characterized by the perimeter of the triangle:  
\be
k \equiv \frac{ k_{1} + k_{2} +k_{3} }{3} \,,
\ee
while the shape is parametrized by the ratios (which are related to the internal angles in the triangle):
\be
\alpha_{1} \equiv 2 \ \frac{k_{2} - k_{3}}{ 3 k} \,, \,\,\,\, \alpha_{2} \equiv \sqrt{3} \ \frac{k_{2} + k_{3} - k_{1}}{3 k} \,.\label{pshape}
\ee
The inverse transformation for the variables $(k_{1}, k_{2}, k_{3})$ reads 
\be \label{kitokalpha} k_{1} = \frac{3 k}{2}\left(1-\frac{\alpha_{2}}{\sqrt{3}}\right) \,, \;\;\;  k_{2} = \frac{3 k}{4}\left(1 + \alpha_{1} +\frac{\alpha_{2}}{\sqrt{3}}\right) \,, \;\;\;  k_{3} = \frac{3 k}{4}\left(1 - \alpha_{1} + \frac{\alpha_{2}}{\sqrt{3}}\right) \,. \ee
We see from the transformation before that each side is separated as $k_{l}= k \gamma_{l}$, where the $\gamma_{l}$ coefficients depend only on the $\alpha_{i}$ variables, so they characterize the shape of the triangle.  An important and useful conclusion that we can extract from Eq. (\ref{ftofiso}) is that one can separate the isotropic part from the terms carrying the anisotropy dependence which is inside the ${(1+\xi_1)}/{ (1+\chi_1)}$ ratio. Then, analogously to the spectral index in the power spectrum in Eq. (\ref{nps}), we define spectral indices for the NG parameter $f_{\rm NL}$. We can use the variable $k$ to define a spectral index related to the scale, and the angles $\alpha_{1}, \alpha_{2}$ to define spectral indices related to the shape in the following way:   
\be\label{spectral} 
n^{k}_{f_{\rm NL}}  \equiv \frac{d \ln f_{\rm NL} }{d\ln k} \,, \qquad n^{\alpha_{1}}_{f_{\rm NL}} \equiv \frac{d \ln f_{\rm NL} }{d\ln \alpha_{1}} \,, \qquad {\rm and} \qquad n^{\alpha_{2}}_{f_{\rm NL}} \equiv \frac{d \ln f_{\rm NL} }{d\ln \alpha_{2}} \,.
\ee
By means of these expressions, we can  evaluate the scale and shape dependence of the NG parameter for general variants of the size and the shape of the momenta triangle. Notice that our definitions will allow us to understand that deviations from scale and shape invariance in $\fnl$ are interpreted as deviations from $n_{\fnl}^k = 0$, $n_{\fnl}^{\alpha_1} = 0$, and $n_{\fnl}^{\alpha_2} = 0$. 
\subsubsection{Scale dependence}
In particular, using Eq. (\ref{ftofiso}) we derive an expression for the spectral index measuring the scale invariance deviation:
\be\label{spectral-k}
n^{k}_{f_{\rm NL}} = n^{k({\rm iso})}_{f_{\rm NL}}  + \frac{1}{1+\xi_{1}} \frac{d \xi_{1}}{d\ln k} - \frac{1}{1+\chi_{1}} \frac{d \chi_{1}}{d\ln k} \,,
\ee
where
\be
n^{k({\rm iso})}_{f_{\rm NL}}  \equiv \frac{d\ln f^{\rm iso}_{\rm NL} }{d \ln k} \,.
\ee
As we can see, in close analogy to the spectral index in Eq. (\ref{nps}), the expression in Eq. (\ref{spectral-k}) allows us to separate the scale dependence of $f_{\rm NL}$ in terms of the scale dependence related to the isotropic part in the first term on the right-hand side of Eq. (\ref{spectral-k}) and the remaining terms that include the contributions related  to the statistical anisotropy dependent terms. In terms of the spectra of the scalar and vector perturbations and their running, the expression in Eq. (\ref{spectral-k}) reads
{\small\ba\label{spectral-k-full}
n^{k}_{f_{\rm NL}} &=& n^{k({\rm iso})}_{f_{\rm NL}}  - \frac{\xi_{1}}{1+\xi_{1}} \left( \frac{d \ln B^{\rm iso}_{\zeta} (k_{1}, k_{2}, k_{3})}{d\ln k} - \frac{d \ln (N_{++}/N^{2}_{+}) }{d\ln k} \right)  \\ \nonumber
&+& \frac{1}{B_{\zeta}} \frac{N_{++}}{N_{+}^{2}}  \sum\limits_{l<m}\left[ P_{\zeta_{A_{+}}}(k_{m}) P^{\rm iso}_{\zeta }(k_{l}) g_{\zeta} (k_{l})\left(\bar{n}_{+}(k_{m}) + \bar{n}_{\zeta}^{\rm iso} (k_{l})+ n_{g}(k_{l}) - 9 \right) (\hat n\cdot \hat k_{l} )^{2} \right. \\ \nonumber
&+&  \left.  P_{\zeta_{A_{+}}}(k_{l}) P^{\rm iso}_{\zeta }(k_{m}) g_{\zeta} (k_{m})\left(\bar{n}_{+}(k_{l}) + \bar{n}_{\zeta}^{\rm iso} (k_{m})+ n_{g}(k_{m}) - 9 \right) (\hat n\cdot \hat k_{m} )^{2} \right. \\ \nonumber
&+&  \left.  P^{\rm iso}_{\zeta }(k_{l}) P^{\rm iso}_{\zeta }(k_{m}) g_{\zeta} (k_{l})g_{\zeta} (k_{m}) \left( \bar{n}_{\zeta}^{\rm iso} (k_{l}) + \bar{n}_{\zeta}^{\rm iso} (k_{m}) + n_{g}(k_{l}) + n_{g}(k_{m}) - 10 \right) (\hat n\cdot \hat k_{l} )(\hat n\cdot \hat k_{m} )(\hat k_{l}\cdot \hat k_{m} )  \right]  \\ \nonumber
&+&  \frac{\chi_{1}}{1+\chi_{1}} \frac{d\ln \left( \sum\limits_{l<m} P_{\zeta}^{\rm iso}(k_{l}) P_{\zeta}^{\rm iso}(k_{m})  \right) }{d\ln k} \\ \nonumber 
&-& \frac{1}{\sum\limits_{l<m}P_{\zeta}(\vec{k}_{l})P_{\zeta}(\vec{k}_{m})}   \sum\limits_{l<m}\left[ P^{\rm iso}_{\zeta }(k_{m}) P^{\rm iso}_{\zeta }(k_{l}) g_{\zeta} (k_{l})\left(\bar{n}_{\zeta}^{\rm iso}(k_{m}) + \bar{n}_{\zeta}^{\rm iso} (k_{l})+ n_{g}(k_{l}) - 9 \right) (\hat n\cdot \hat k_{l} )^{2} \right. \\ \nonumber
&+&  \left.  P^{\rm iso}_{\zeta }(k_{l}) P^{\rm iso}_{\zeta }(k_{m}) g_{\zeta} (k_{m})\left(\bar{n}_{\zeta}^{\rm iso}(k_{l}) + \bar{n}_{\zeta}^{\rm iso} (k_{m})+ n_{g}(k_{m}) - 9 \right) (\hat n\cdot \hat k_{m} )^{2} \right. \\ \nonumber
&+&  \left.  P^{\rm iso}_{\zeta }(k_{l}) P^{\rm iso}_{\zeta }(k_{m}) g_{\zeta} (k_{l})g_{\zeta} (k_{m}) \left( \bar{n}_{\zeta}^{\rm iso} (k_{l}) + \bar{n}_{\zeta}^{\rm iso} (k_{m}) + n_{g}(k_{l}) + n_{g}(k_{m}) - 10 \right) (\hat n\cdot \hat k_{l} )^{2}(\hat n\cdot \hat k_{m} )^{2}  \right] . \nonumber
\ea }
To obtain the latter formula, we have introduced $\bar{n}= n+3$ for all the spectral indices and used the following expression, which is valid for any power spectra $P(k)$:
\be k\frac{d P(k_{l})}{dk} = k_{l} \frac{d P(k_{l})}{dk_{l}} = P(k_{l})(\bar{n}(k_{l}) - 4) \,.\ee

In order to estimate the deviation from scale invariance in $f_{\rm NL}$ caused by the presence of statistical anisotropy, let us consider an equilateral configuration so that all the spectral indices $n_{+}, n^{\rm iso}_{\zeta}$,  and $n_{\phi}$ depend on the same scale. This configuration can be achieved when $\alpha_{1} = 0$ and $\alpha_{2} = \sqrt{3}/3$, which implies $k_{i} = k$. We use as a concrete example a scalar field and a single vector field with a constant ratio $q = P_{{\rm long}}/ P_{+}$. We recall that, in this case, we have $n_{g}-1 = n_{+} - n_{\zeta}^{\rm iso}$ and, therefore, we get that Eq. (\ref{spectral-k-full}) reduces to
\ba\label{spectral-k-massless} 
n^{k}_{f_{\rm NL}}  &=&  n^{k({\rm iso})}_{f_{\rm NL}}  + \frac{ \xi_{1}}{ (1+  \xi_{1})   } \left[ 2( \bar{n}_{+} - 4 ) + \frac{d \ln   ({N_{++}/N^{2}_{+}})}{d\ln k}   \right. \\ \nonumber
&-& \left( 2(\bar{n}_{+} - 4) + \frac{d \ln (N_{++}/N^{2}_{+}) }{d\ln k}  \right)  \frac{N_{++}}{ B^{\rm iso}_{\zeta} N^{2}_{+}} 3 \left( P_{\zeta_{A_{+}}} (k)\right)^{2}   \\ \nonumber
&-& \left.  \left( 2(\bar{n}_{\phi} - 4) + \frac{d \ln (N_{\phi \phi}/N^{2}_{\phi}) }{d\ln k}  \right) \frac{N_{\phi \phi}}{ B^{\rm iso}_{\zeta} N^{2}_{\phi}} 3 \left( P_{\zeta_{A_{\phi}}} (k)\right)^{2}\right] \\ \nonumber
&-&  \frac{\chi_{1}}{1+\chi_{1}} \left(   n_{g} -1  \right) \left( 1 -   { \left(P^{\rm iso}_{\zeta }(k)  g_{\zeta} (k) \right)^{2} \sum\limits_{l<m}  (\hat n\cdot \hat k_{l} )^{2}  (\hat n\cdot \hat k_{m} )^{2} }\right).
\ea 
In this situation, we have
\ba\label{spectral-k-iso} 
n^{k({\rm iso})}_{f_{\rm NL}} &=&  -2 (\bar{n}_{\zeta}^{\rm iso} - 4 )  \\ \nonumber
&+& \left( 2(\bar{n}_{+} - 4) + \frac{d \ln (N_{++}/N^{2}_{+}) }{d\ln k}  \right)  \frac{N_{++}}{ B^{\rm iso}_{\zeta} N^{2}_{+}} 3 \left( P_{\zeta_{A_{+}}} (k)\right)^{2}   \\ \nonumber
&+& \left( 2(\bar{n}_{\phi} - 4) + \frac{d \ln (N_{\phi \phi}/N^{2}_{\phi}) }{d\ln k}  \right)  \frac{N_{\phi \phi}}{ B^{\rm iso}_{\zeta} N^{2}_{\phi}} 3 \left( P_{\zeta_{A_{\phi}}} (k)\right)^{2}.  
\ea 
We can test the obtained expressions by going to some well-known limiting cases, for instance, a single inflaton field;  in such a case, we obtain
\be
n^{k}_{f_{\rm NL}}  =  n^{k({\rm iso})}_{f_{\rm NL}} =  \frac{d \ln (N_{\phi \phi}/N^{2}_{\phi}) }{d\ln k} \,,
\ee 
which is the known result for single field inflation and is a first order quantity in the slow-roll parameters \cite{Byrnes09}.  In the same way, if vector perturbations dominate over scalar perturbations, the result is
\be
n^{k}_{f_{\rm NL}}  =  n^{k({\rm iso})}_{f_{\rm NL}} =  \frac{d \ln (N_{++}/N^{2}_{+}) }{d\ln k} \,.
\ee 
Scale dependent deviations in the anisotropic terms on the right-hand side of Eq. (\ref{spectral-k-massless}) appear due to the simultaneous presence of scalar and vector perturbations. If there were only scalar or only vector perturbations, the anisotropic terms on the right-hand side of Eq. (\ref{spectral-k-massless}) would be zero. We will evaluate the weight of these deviations for different situations and limits in section \ref{eval}. \\

\subsubsection{Shape dependence}
In the same way, we can evaluate the shape dependence through the spectral indices in the angular parameters $\alpha_{1}$ and $\alpha_{2}$. For simplicity, we can choose $\alpha_{1}=0$ while keeping $\alpha_2$ as a variable which corresponds to an isosceles configuration. Thus, we can easily go to the squeezed limit ($\alpha_{2} = \sqrt{3}$), the equilateral limit ($\alpha_{2} = \sqrt{3}/3$), and the folded limit ($\alpha_{2} = 0$). Then, analogously to the running with the scale, we calculate
\be\label{spectral-alpha}
n^{\alpha_{2}}_{f_{\rm NL}} = n^{\alpha_{2}({\rm iso})}_{f_{\rm NL}}  + \frac{1}{1+\xi_{1}} \frac{d \xi_{1}}{d\ln \alpha_{2}} - \frac{1}{1+\chi_{1}} \frac{d \chi_{1}}{d\ln \alpha_{2}} \,.
\ee
In this case, we have that the logarithmic derivatives of the spectra with respect to the angular variable $\alpha_2$ are calculated by means of the formula 
\be\label{alphapower} \alpha_{2}\frac{d P(k_{l})}{d\alpha_{2}} = \alpha_{2} \frac{d k_{l}}{d\alpha_{2}}  \frac{d P(k_{l})}{d k_{l}} = \alpha_{2} \frac{\gamma_{l} }{k_{l} }P(k_{l})(\bar{n}(k_{l}) - 4) \,,\ee
where the $\gamma_{l}$ constant coefficients are derived directly from the change of coordinates  in Eq. (\ref{kitokalpha}), being $\gamma_{l}=\partial k_{l}/\partial \alpha_{2}$. Besides the dependence of the spectra, we should also evaluate the derivatives of the scalar products $\hat n\cdot \hat k_{l}$ and $\hat k_{l}\cdot \hat k_{m}$, which leads to long analytic expressions for the spectral indices; then, instead of writing the analytic expressions, we will evaluate some of their important limits in section \ref{eval}.

\section{Evaluation of the scale and shape dependence}\label{eval}
\begin{figure}[h!] 
        \centering
        \includegraphics[width=9.5cm]{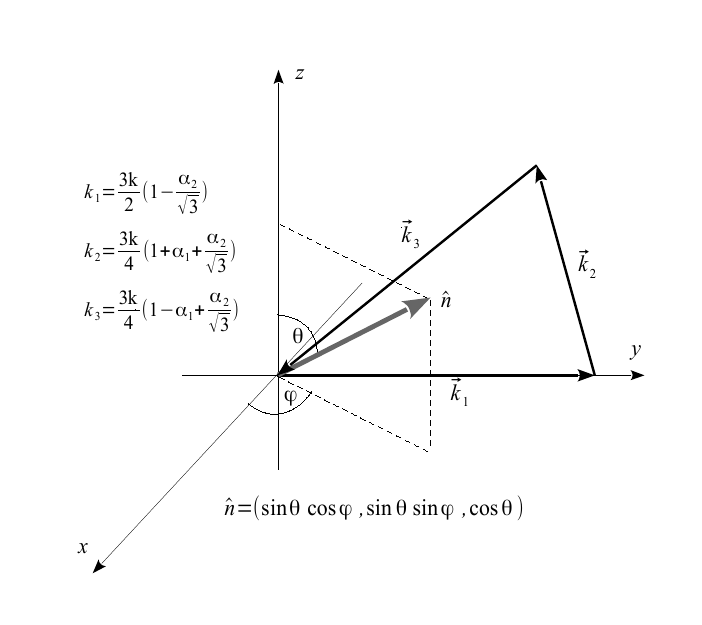}
                \caption{Parameters for the size and shape of the momenta configuration.}
                \label{shape_par} 
\end{figure}

In the following, we will use the parameters represented in Fig. \ref{shape_par}. As shown in the figure, due to the momentum conservation (or equivalently, due to statistical homogeneity), the momentum vectors ${\vec k}_{i}$ are restricted to form a closed triangle and the vector $\hat{n}$ is described by their polar and azimuthal angles $\theta$ and $\phi$, respectively. Then, the BS and $f_{\rm NL}$ are functions of the angles that define the orientation of $\hat{n}$, and of the parameters of the triangle formed by the momentum vectors, i.e., $k, \alpha_{1}$, and $\alpha_{2} $  defined in Eq. (\ref{kitokalpha});  thus, $f_{\rm NL } = f_{\rm NL }(k, \alpha_{1}, \alpha_{2}, \theta, \phi)$. Of course, the expression for $f_{\rm NL}$ depends on the derivatives of $N$ and on the amplitudes and the spectral indices of the scalar and vector perturbations.  In order to perform our evaluations, we will follow the configuration represented in Fig. \ref{shape_par} and specialize to the massless vector case in which $q=0$ and $-1 \leqslant \gz<0$ [see Eq. (\ref{g-anis})]. The $f(\phi)F^{2}$ model of Ref. \cite{Watanabe:2009ct} enters in this category. 
\subsection{$\fnl/\fnl^{\rm iso}$ ratio}
\begin{figure}[h!]
        \centering
        \begin{subfigure}[b]{0.495\textwidth}
                \includegraphics[width=\textwidth]{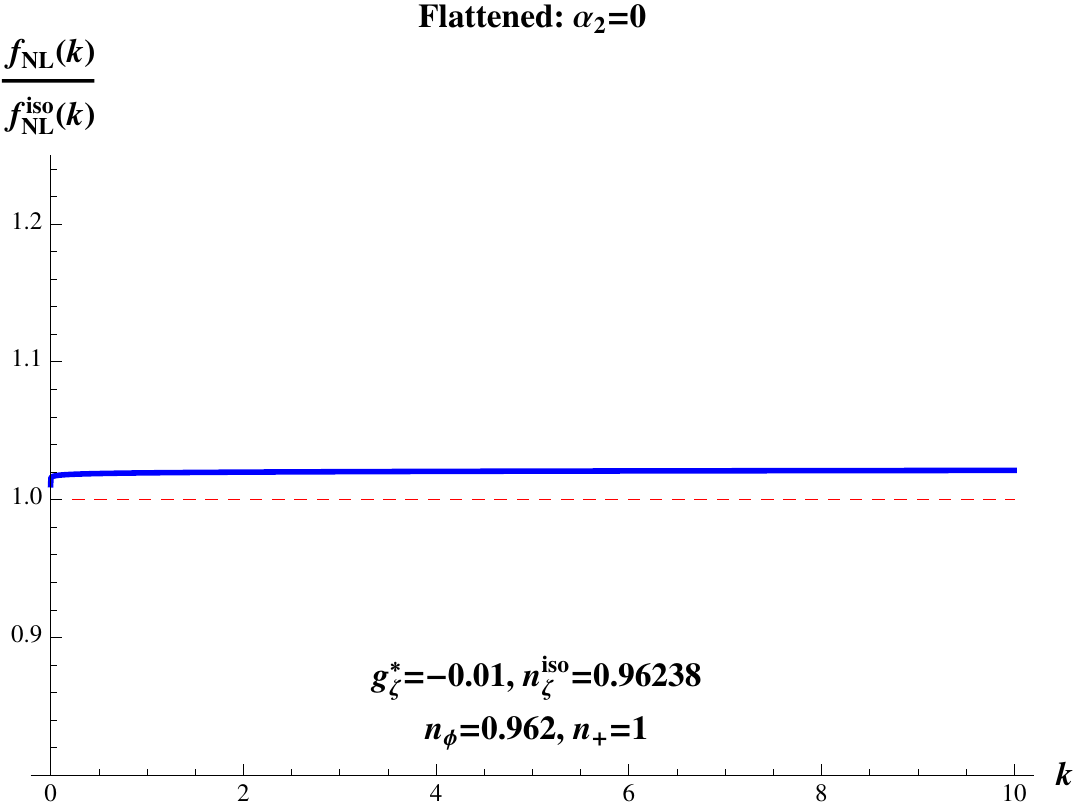}
                \caption{Flattened configuration with $g_{\zeta}^{*}=-0.01$.}
                \label{fg0.01}
        \end{subfigure}
\begin{subfigure}[b]{0.495\textwidth}
                \includegraphics[width=\textwidth]{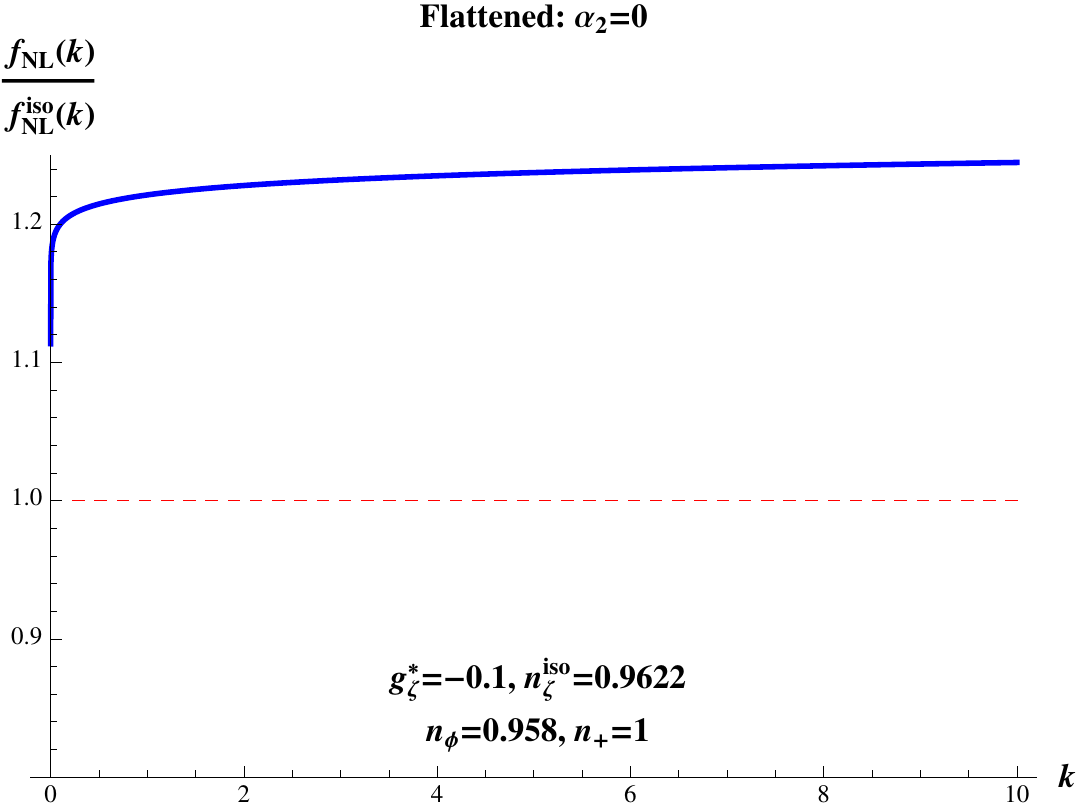}
                \caption{Flattened configuration with $g_{\zeta}^{*}=-0.1$.}
                \label{fg0.1}
        \end{subfigure}        
\caption{Ratio  $\fnl/\fnl^{\rm iso}$  for flattened configurations ($\alpha_{2}=0, k_{1}=k_{2}+k_{3}$) with values (a) $g_{\zeta}^{*}=-0.01$  and (b) $g_{\zeta}^{*}=-0.1$. We choose the spectral indices such that  vector perturbations are scale invariant, $n_{+}=1$, and for scalar perturbations we set $n_{\phi}= 0.962$ for  $g_{\zeta}^{*}=-0.01$ and $n_{\phi}= 0.958$ for  $g_{\zeta}^{*}=-0.1$ so that the spectral index of the primordial curvature perturbation is close to the current observed value $n_{\zeta}=0.962$ \cite{Bennett12,PlanckCP13}. In both plots, we have set $\theta = \varphi=\pi/2$ because the ratio is maximized at these values, i.e., when $\nn$ coincides with $\kn_{1}$. We see that, for this configuration and for  $g_{\zeta}^{*}=-0.01$, $\fnl$ is enhanced around 2$\%$, while for   $g_{\zeta}^{*}=-0.1$ the enhancement is around 20$\%$ with respect to the isotropic value.  }\label{flat}
\end{figure}

\begin{figure}[h!]
        \centering
        \begin{subfigure}[b]{0.495\textwidth}
                \includegraphics[width=\textwidth]{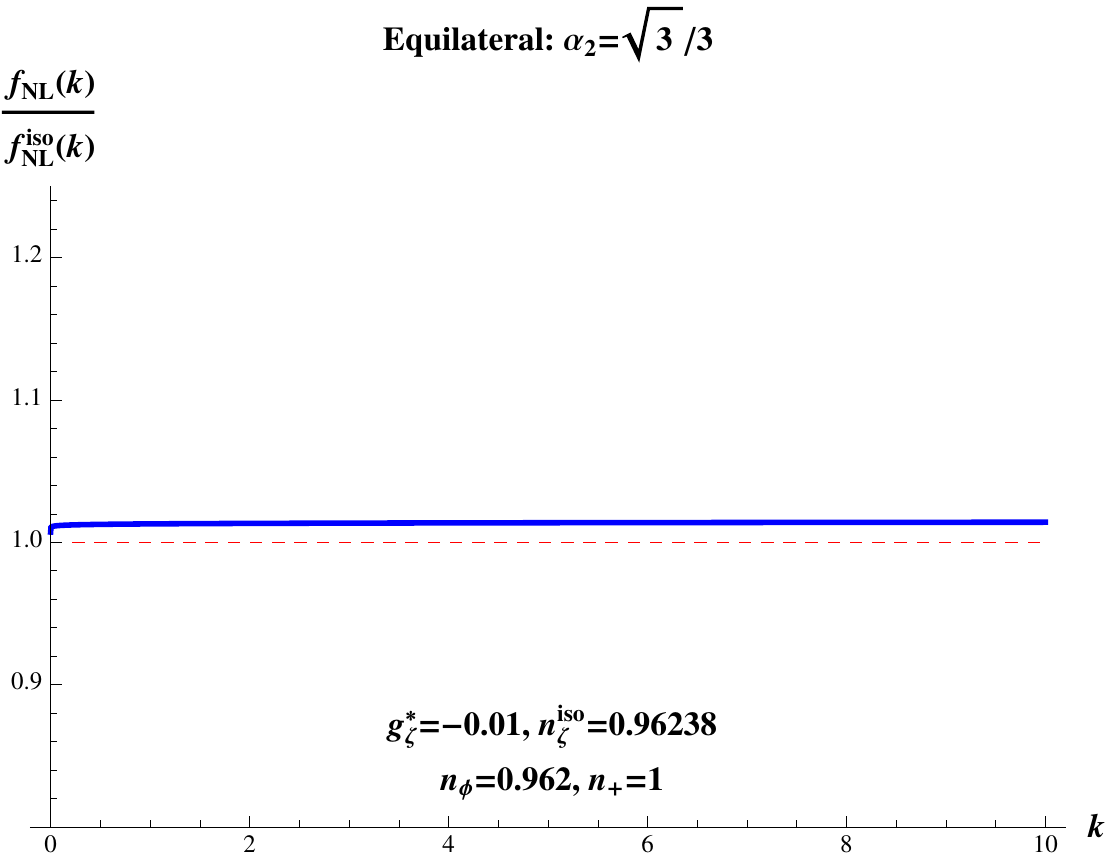}
                \caption{Equilateral configuration with $g_{\zeta}^{*}=-0.01$.}
                \label{eq0.01}
        \end{subfigure}
\begin{subfigure}[b]{0.495\textwidth}
                \includegraphics[width=\textwidth]{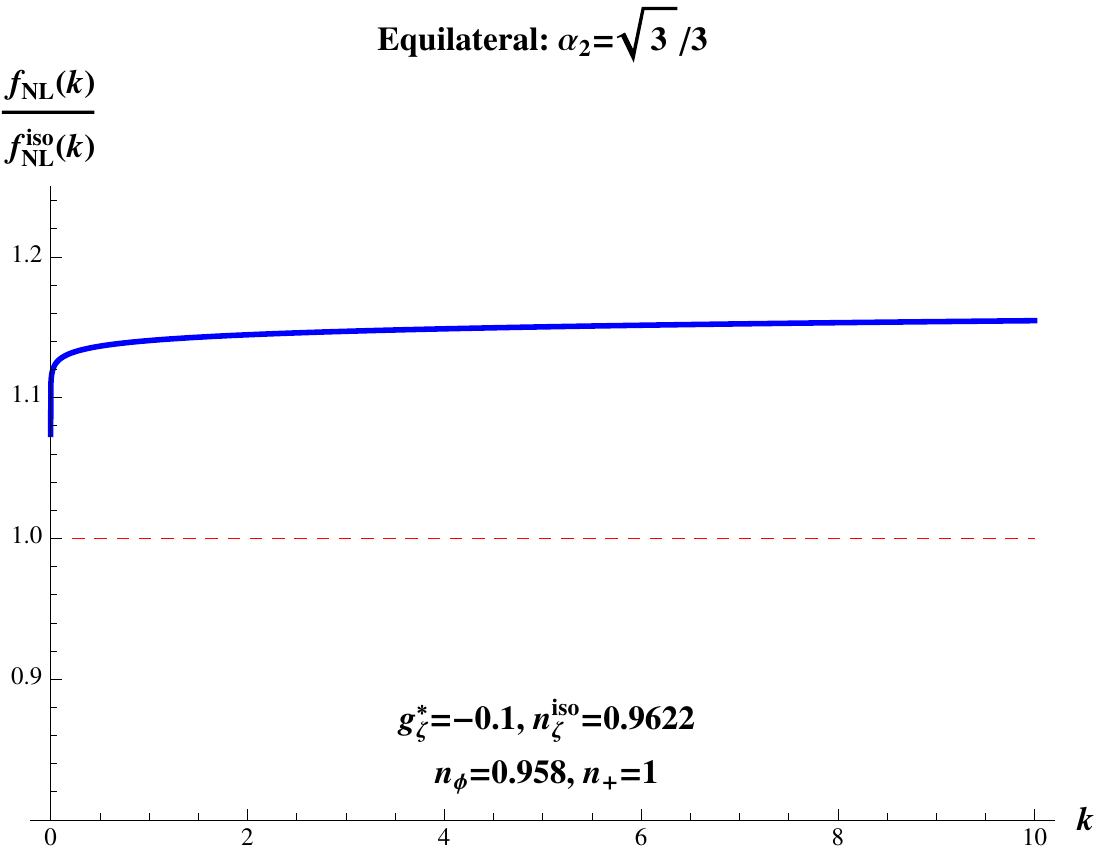}
                \caption{Equilateral configuration with $g_{\zeta}^{*}=-0.1$.}
                \label{eq0.1}
        \end{subfigure}        \caption{The same as in Fig. \ref{flat} but for the equilateral configurations. In this case, the ratio is suppressed so that the enhancement of the $\fnl$ parameter is about 1$\%$ for (a) $g_{\zeta}^{*}=-0.01$ and over 10$\%$ for (b) $g_{\zeta}^{*}=-0.1$.}\label{equi}
\end{figure}

\begin{figure}[h!]
        \centering
        \begin{subfigure}[b]{0.495\textwidth}
                \includegraphics[width=\textwidth]{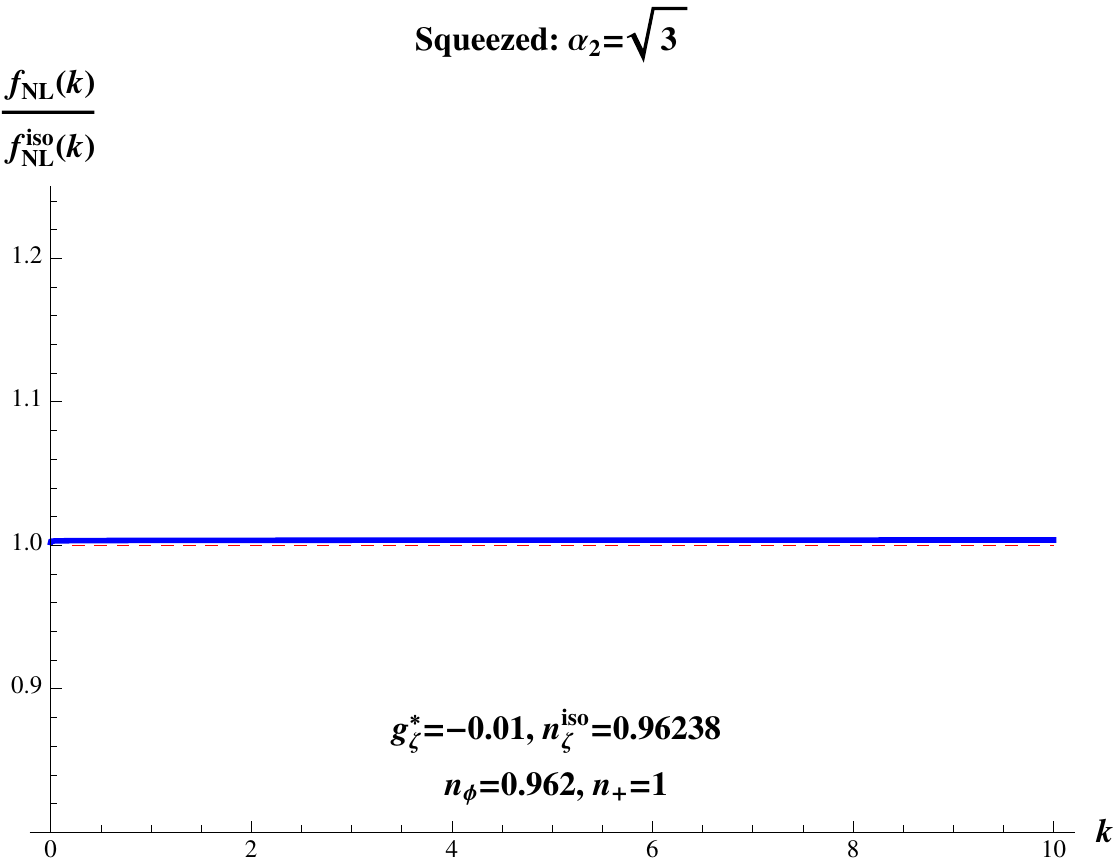}
                \caption{Squeezed configuration with $g_{\zeta}^{*}=-0.01$.}
                \label{sqz0.01}
        \end{subfigure}
\begin{subfigure}[b]{0.495\textwidth}
                \includegraphics[width=\textwidth]{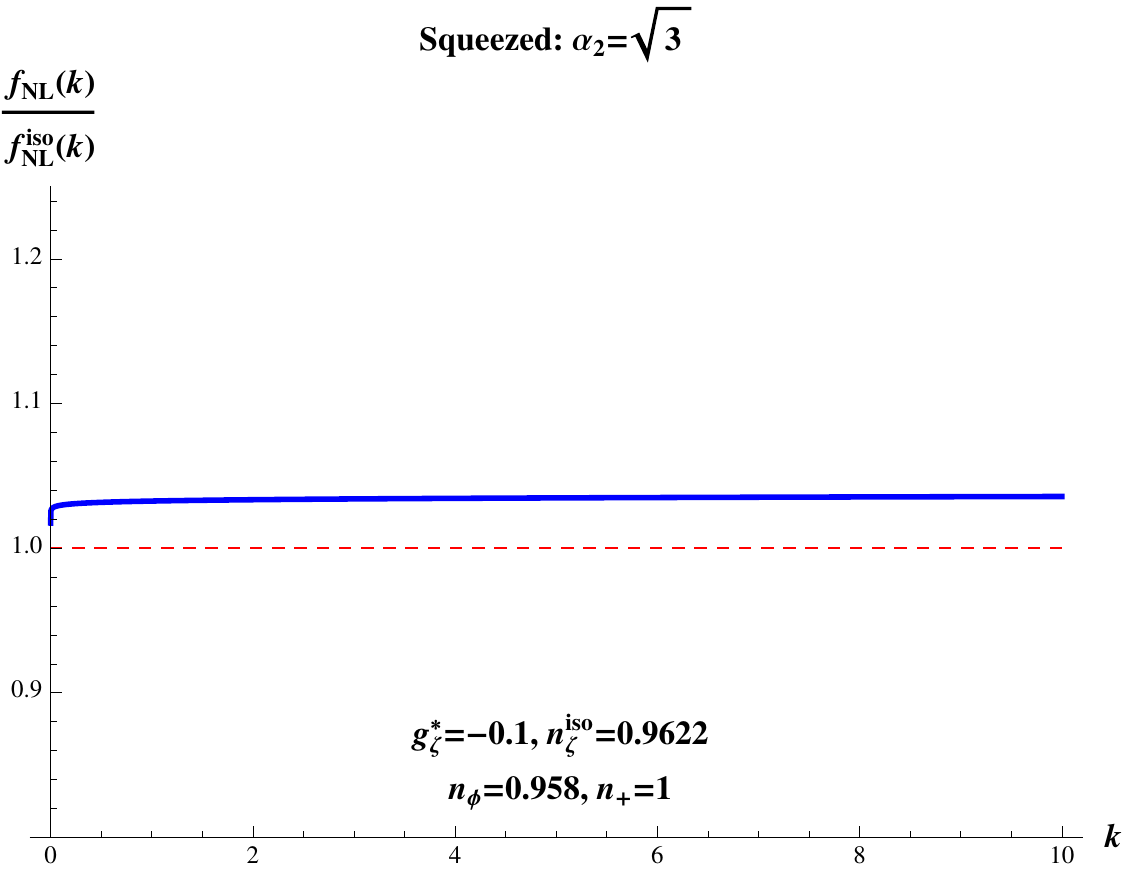}
                \caption{Squeezed configuration with $g_{\zeta}^{*}=-0.1$.}
                \label{sqz0.1}
        \end{subfigure}        \caption{The same as in Fig. \ref{flat} but for the squeezed configurations. In this case, the ratio is even more suppressed so that the ratio is around 0.1 $\%$ for (a) $g_{\zeta}^{*}=-0.01$ and  around 2$\%$ for  (b) $g_{\zeta}^{*}=-0.1$.}\label{squeezed}
\end{figure}

To estimate the contribution of the anisotropic terms to the non-Gaussianity parameter $\fnl$, which is present in Eq. (\ref{ftofiso}), we first evaluate the dependence of the ratio $\fnl/\fnl^{\rm iso}$ on the size and shape of the momenta triangle from the expression in Eq. (\ref{ftofiso}). In Figs. \ref{flat}, \ref{equi} and \ref{squeezed} we show the results of the evaluation of  Eq. (\ref{ftofiso})  in the flattened (or folded), equilateral,  and squeezed configurations, respectively. We see that the deviations from the value in the isotropic case are significant for specific configurations of the $\nn$ and the $\kn_{i}$. In all the plots in Figs. \ref{flat}, \ref{equi} and \ref{squeezed}, we use $\theta=\varphi=\pi/2$. For this set of parameters, which represents the case where the vector $\nn$ coincides with the vector $\kn_{1}$, the effects of the anisotropic terms are maximized. The $\fnl$ parameter and the ratio $\fnl/\fnl^{\rm iso}$ are sensitive to many details, mainly to  the spectral indices of the scalar and vector perturbations and the ratios of the amplitudes ${\cal A}_{ +}$ and ${\cal A}_{ \phi}$.   We choose $\gz^{*}=-0.01$ (at the pivot scale $k=k^{*}$) for the appropriate ${\cal A}_{ +}/{\cal A}_{ \phi}$ ratio, which is still within the margin allowed by Planck, once the effects of the asymmetric beams have been removed  \cite{Kim13}. We also choose an almost scale invariant spectrum so that we set the spectral indices of the scalar perturbations close to the current observational value for the spectral index of the primordial curvature perturbation $n_{\zeta}^{\rm iso} \approx 0.962$.  This results in an enhancement of the $\fnl$ parameter, which for all the configurations is around the $0.1$ to 1 percent-level order. We also plot the deviations with $\gz^{*}=-0.1$. For this case, we get an enhancement of over 20$\%$ for the flattened configuration in which we obtain the maximum effect. For equilateral and squeezed configurations, we obtain around 10$\%$ and 2$\%$ enhancements, respectively. Certainly, $\gz^{*}=-0.1$ is excluded by current observations, and our intention here with this evaluation is to test the sensitivity of the $\fnl/\fnl^{\rm iso}$ quotient with respect to $\gz^{*}$. As a result, we get that $\gz^{*}=-0.1$ produces a significant enhancement of $\fnl$ inducing a significant level of anisotropic NG. This enhancement could be important if $\fnliso$ is big enough to be detected; otherwise, even a 20$\%$ enhancement would turn out to be difficult to discern from the full signal. In order to do a precise statement about the detectability of the anisotropic signal, it would be important to use the {\it primordial shape correlator}  defined in \cite{Fergusson08}. Such a correlator allows us to determine the independence of different shapes of non-Gaussianity in terms of the bispectrum. In our case, we should evaluate the correlation of the isotropic BS with respect to the anisotropic BS, using Eq. (\ref{bs0}). Nevertheless, a crude estimate of the correlation between the isotropic and the anisotropic BS tells us that both forms are closely correlated, the level of independence being the order of the statistical anisotropy parameter $\gz^{*}$. Even for a very high value such as $\gz^{*}=-0.1$, we would need a very high precision, several $\sigma$ detection, in order to detect an anisotropic signal. Certainly, for more realistic values, such as $\gz^{*}=-0.01$, the detection of an anisotropic signal seems to be practically impossible to achieve.  

The $\fnl$ parameter  is also sensitive to the ratio of the fractions $(N_{++}/N_{+}^{2})$ and $(N_{\phi \phi}/N_{\phi}^{2})$,  but this sensitivity is negligible unless the difference between both fractions is such that $(N_{++}/N_{+}^{2}) \gg N_{\phi \phi}/N_{\phi}^{2}$ (a difference that is at least bigger than 2 orders of magnitude) and for vector perturbations with spectra far from being scale invariant. This situation can be
realized if we had strongly scale dependent vector perturbations which strongly dominated over scalar perturbations in the bispectrum [see Eq. (\ref{bs_vec+scalar})], which does not seem to be too realistic; as a result, in such a situation, we would obtain large values of anisotropic non-Gaussianity which, being above the observed limits for the isotropic case, are hardly credible, although there are actually no data analyses of non-Gaussianity involving anisotropy (however, see Ref. \cite{Bartolo11});   for this reason, we will not consider this case here. If some phenomenological or observational evidence appeared that supported the presence of strongly scale dependent vector perturbations dominating the bispectrum, it would be interesting to consider this case seriously. In the meantime, we will keep both fractions within the same order of magnitude and, as a result, the $\fnl/\fnl^{\rm iso}$ ratio will remain stable, allowing us to obtain practically the same results for all the configurations studied. 

\begin{figure}[h!]
        \centering
        \begin{subfigure}[b]{0.495\textwidth}
                \includegraphics[width=\textwidth]{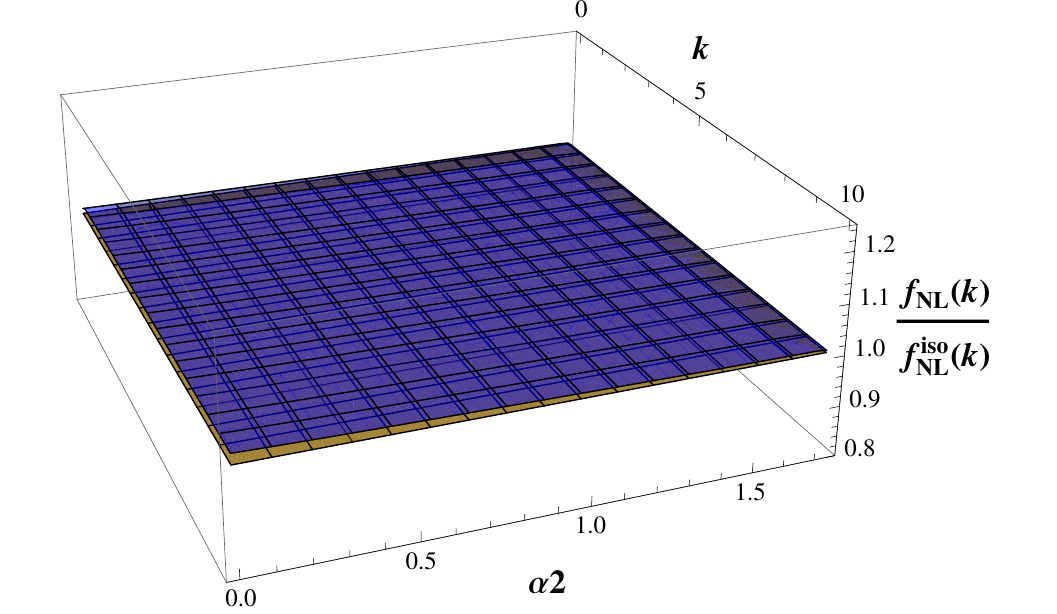}
                \caption{$\fnl / \fnl^{\rm iso}$ ratio with $g_{\zeta}^{*}=-0.01$.}
                \label{fnlg0.01}
        \end{subfigure}
\begin{subfigure}[b]{0.495\textwidth}
                \includegraphics[width=\textwidth]{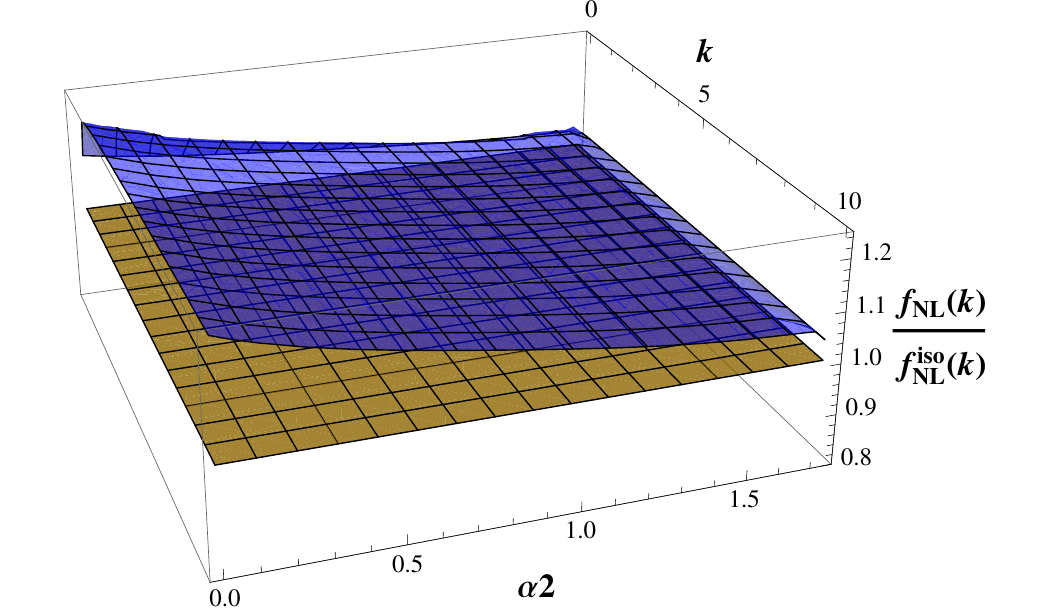}
                \caption{$\fnl / \fnl^{\rm iso}$ ratio with $g_{\zeta}^{*}=-0.1$.}
                \label{fnlg0.1}
        \end{subfigure}        
\caption{Ratio  $\fnl/\fnl^{\rm iso}$ with (a) $g_{\zeta}^{*}=-0.01$ and (b) $g_{\zeta}^{*}=-0.1$  for all the configurations described by changing the shape parameter $\alpha_{2}$. }\label{fnlgplot}
\end{figure}

To summarize,  in Fig. \ref{fnlgplot} we plot the $\fnl/\fnl^{\rm iso}$ ratio for all the values of the shape parameter $\alpha_{2}$ with both $g_{\zeta}^{*}=-0.01$ and $g_{\zeta}^{*}=-0.1$. 
\subsection{Scale dependence}
\begin{figure}[h!]
        \centering
        \begin{subfigure}[b]{0.495\textwidth}
                \includegraphics[width=\textwidth]{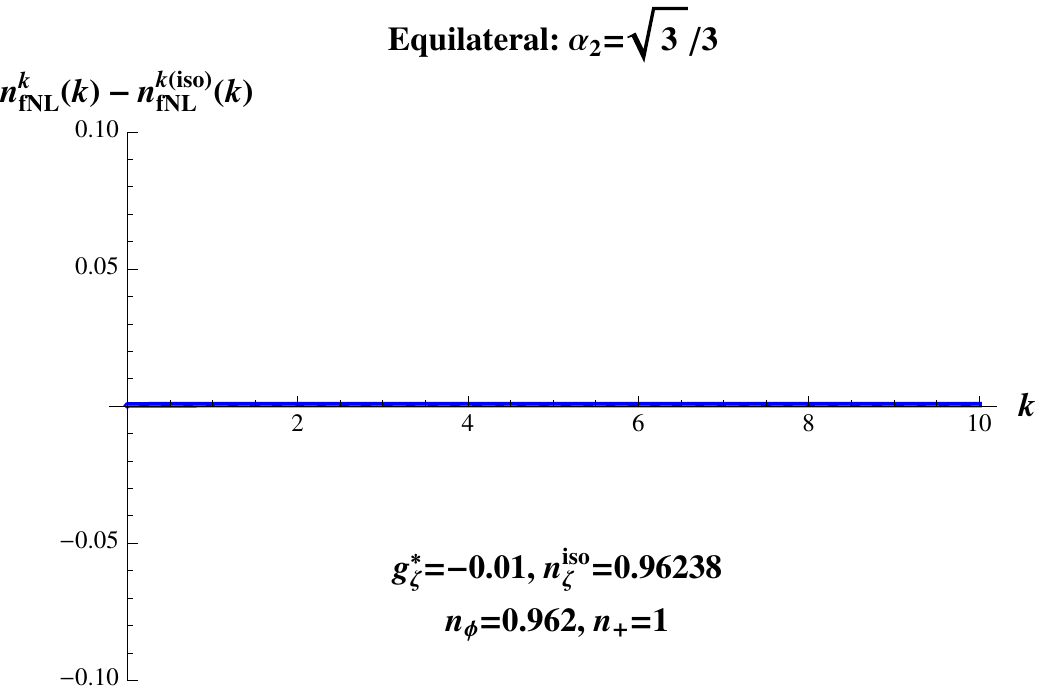}
                \caption{$n_{f_{\rm NL}}^{k} - n_{f_{\rm NL}}^{k({\rm iso})}$  with $g_{\zeta}^{*}=-0.01$.}
                \label{nkg001}
        \end{subfigure}
\begin{subfigure}[b]{0.495\textwidth}
                \includegraphics[width=\textwidth]{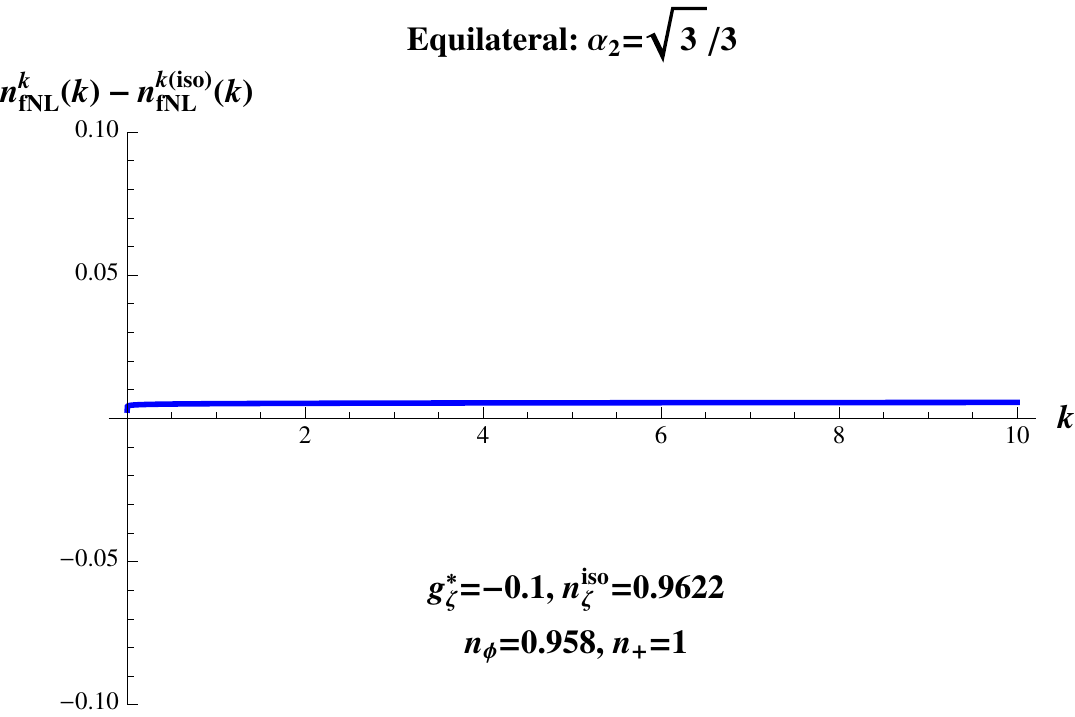}
                \caption{$n_{f_{\rm NL}}^{k} - n_{f_{\rm NL}}^{k({\rm iso})}$  with $g_{\zeta}^{*}=-0.1$.}
                \label{nkg01}
        \end{subfigure}        
\caption{Deviation of the spectral index $n_{f_{\rm NL}}^{k}$ with respect to the isotropic case with (a) $g_{\zeta}^{*}=-0.01$ and (b) $g_{\zeta}^{*}=-0.1$ for equilateral configurations.}\label{nfnlgk}
\end{figure}

Now, we plot the results of the deviations from scale invariance of the $\fnl$ parameter derived from the expression in Eq. (\ref{spectral-k-massless}).
In Fig. \ref{nfnlgk} we see the evaluation for the deviation of the spectral index $n^k_{f_{\rm NL}}$ with respect to the isotropic case: $\Delta n_{f_{\rm NL}}^{k}  \equiv n_{f_{\rm NL}}^{k} - n_{f_{\rm NL}}^{k({\rm iso})}$ derived from Eq. (\ref{spectral-k-massless}) in the equilateral limit. In this case, we perceive that the deviation from the isotropic case is modified by the terms on the right-hand side of  Eq. (\ref{spectral-k-massless}) which are strongly dependent on the configuration of the $\kn$ and $\nn$ vectors. The deviation is suppressed by the statistical anisotropy level $\gz$ and the slow-roll parameters within the spectral indices $n_{+}, n_{\phi}$, and $n_{g}$. We neglect the second order contributions in slow roll coming from the scale derivatives of the $N_{++}/N_{+}^{2}$ and $N_{\phi \phi}/N_{\phi}^{2}$ terms. As we can perceive, the modifications of the scale dependence are negligible for scale invariant vector perturbations and for $-0.1<\gz^{*} < -0.01$. Again, as in the evaluation of the $\fnl/\fnl^{\rm iso}$ ratio, the spectral index would acquire significant corrections if we went to configurations with strongly scale dependent vector perturbations and high values of the statistical anisotropy parameters, and if vector perturbations strongly dominated the bispectrum of the primordial curvature perturbation. We will not exhibit these cases here. 
\subsection{Shape dependence}
The spectral index for the shape dependence described by the angular parameter  $\alpha_{2}$ is more complicated and, at the same time, is potentially more interesting due to the intricate structure of the logarithmic derivatives of the scalar products $(\kn_{l}\cdot\nn)$ and $(\kn_{l}\cdot \kn_{m})$. For instance, the squeezed limit ($\alpha_{2}=\sqrt{3}$) is complicated to work with since the $\gamma_{1}/k_{1}$ coefficient in Eq. (\ref{alphapower}) is singular at $\alpha_{2}=\sqrt{3}$. There are more configurations which greatly enhance the isotropic index  even for $g_{\zeta}^{*}\approx-0.01$ and a nearly scale invariant spectrum of scalar and vector perturbations.

\begin{figure}[h!]
        \centering
        \begin{subfigure}[b]{0.495\textwidth}
                \includegraphics[width=\textwidth]{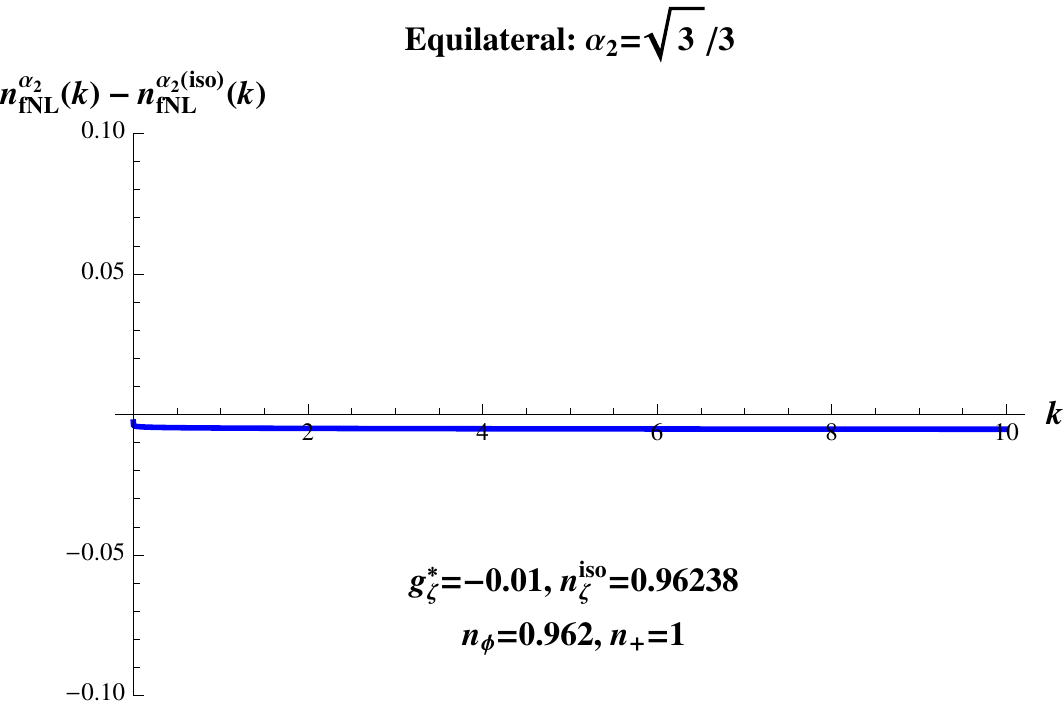}
                \caption{$n_{f_{\rm NL}}^{\alpha_{2}} - n_{f_{\rm NL}}^{\alpha_{2}({\rm iso})}$  with $g_{\zeta}^{*}=-0.01$.}
                \label{nkal001}
        \end{subfigure}
\begin{subfigure}[b]{0.495\textwidth}
                \includegraphics[width=\textwidth]{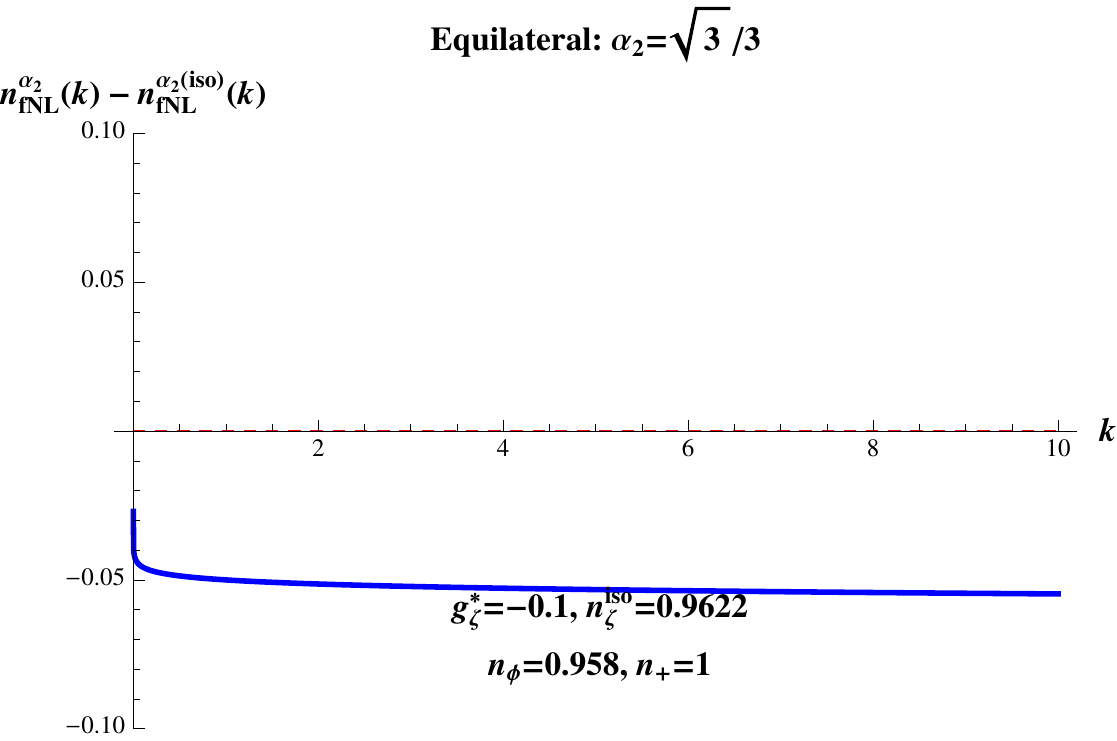}
                \caption{$n_{f_{\rm NL}}^{\alpha_{2}} - n_{f_{\rm NL}}^{\alpha_{2}({\rm iso})}$  with $g_{\zeta}^{*}=-0.1$.}
                \label{nkal01}
        \end{subfigure}        
\caption{Deviation of the spectral index $n_{f_{\rm NL}}^{\alpha_{2}}$ with respect to the isotropic case with (a) $g_{\zeta}^{*}=-0.01$ and (b) $g_{\zeta}^{*}=-0.1$ for equilateral configurations.}\label{nalpha}
\end{figure}

\begin{figure}[h!]
        \centering
                \includegraphics[width=0.495\textwidth]{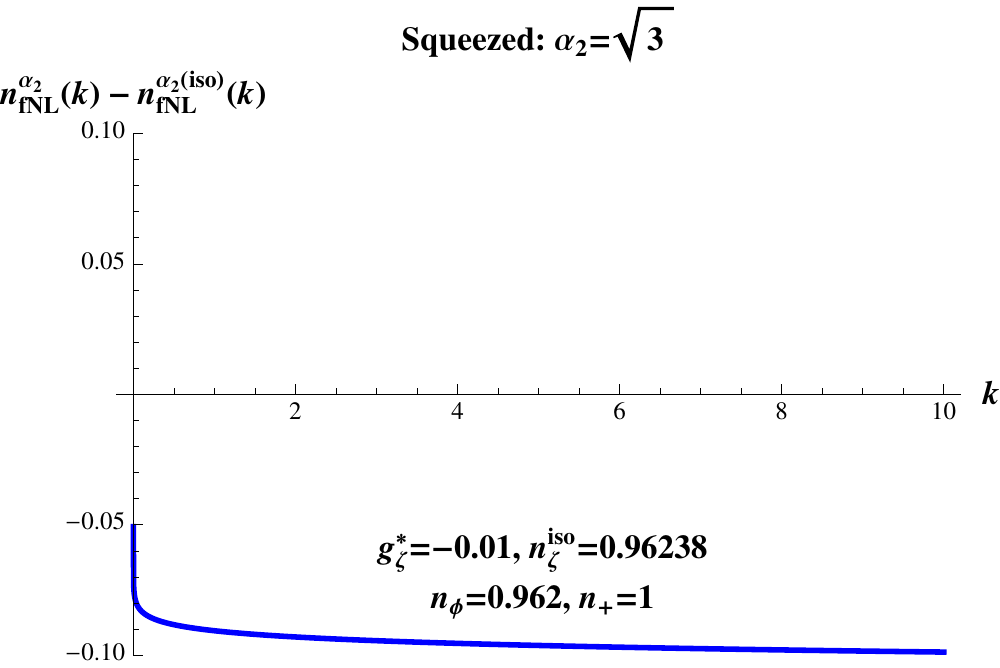}
\caption{Deviation of the spectral index $n_{f_{\rm NL}}^{\alpha_{2}}$ with respect to the isotropic case with $g_{\zeta}^{*}=-0.01$ for squeezed configurations.}\label{nalphaSqz}
\end{figure}

In Fig. \ref{nalpha}, we see the evaluation of $n^{\alpha_2}_{f_{\rm NL}} - n^{\alpha_2 (\rm iso)}_{f_{\rm NL}}$, coming from Eq. (\ref{spectral-alpha}), in equilateral configurations for the same values of the level of statistical anisotropy and the spectral indices used in the evaluations of $\fnl/\fnl^{\rm iso}$ and $n_{f_{\rm NL}}^{k} - n_{f_{\rm NL}}^{k({\rm iso})}$, whereas, in Fig. {\ref{nalphaSqz}}, we go to the squeezed configuration with  $g_{\zeta}^{*}=-0.01$. As said before, in the latter limit, the modification is noticeable and gives $\Delta n_{f_{\rm NL}}^{\alpha_{2}} \approx -0.1$ even though the level of statistical anisotropy is small. 

To conclude this section, we would like to point out that there are several generalizations and variants to the calculations we did here. For instance, it would be instructive, and rather straightforward, to generalize our results to include multiple scalars and vector fields. It would also be interesting to consider a general curved metric in the field space as discussed in Ref. \cite{Byrnes12a}, where the authors show that it induces scale dependent effects in the correlators, in particular, in the $f_{\rm NL}$ parameter. Our results here constitute another source of scale and shape dependence besides those considered in that reference. Another approach regarding scale dependent anisotropic NG without invoking vector fields, but using instead noncommutativity of space-time to generate anisotropic correlators, was discussed in Ref. \cite{Nautiyal2013}. Finally, it would also be very interesting to study the parametrization of the BS presented in Ref. \cite{Shiraishi13} in terms of a series of Legendre polynomials $P_{L}$: 
\be
B_{\zeta}({\vec k}_1 ,\,{\vec k}_2 ,\, {\vec k}_3 ) = \sum_{L}  \sum\limits_{l<m}  c_{L} P_{L}(\hat k_{l} \cdot \hat k_{m}) P_{\zeta} (k_{l}) P_{\zeta} (k_{m}) \,,
\ee
where the $c_{L}$'s are constant weight coefficients which depend on the particular model. It has been discussed in Ref. \cite{Shiraishi13} that the $c_{L}$ coefficients are sensitive to the presence of vector fields, and it was shown that models in which $\zeta$ is sourced by the anisotropic stress of large-scale magnetic fields, models with an $f(\phi) F^{2}$ interacting term (see, e.g., Ref. \cite{Bartolo12} and  references therein), and the ``{\it solid inflation}'' model \cite{Endlich12, Bartolo13} are particular examples of this parametrization producing specific nonzero values for the $c_{L}$ coefficients. The analysis done in Ref. \cite{Shiraishi13} assumes the $c_{L}$'s are scale independent, making it interesting to consider models with solid motivations for the introduction of scale dependent effects. 
 
\section{Results and discussion}\label{results}
The results obtained here allow us to conclude that the presence of vector fields during inflation introduces several interesting and appealing features, not only for fundamental theoretical reasons, but also from the quantitative point of view. Their presence offers us new challenges for the precise interpretation and, hopefully, the measurement of observables related to the presence of inflationary vector fields. The first and most evident manifestation of vector fields appears in the form of statistical anisotropy in the correlation functions of the primordial curvature perturbation. Here, we consider models which exhibit statistical anisotropy in the power spectrum of a quadrupolar form which allows us to quantify the level of statistical anisotropy by the single parameter $\gz.$ With this starting point, we study the consequences of this type of statistical anisotropy in the higher order correlators, particularly in the bispectrum and in the NG parameter $\fnl$.  

We first take into account that, generically, these models introduce a level of scale dependence in the $\gz$ parameter if the spectral indices of scalar and vector perturbations are different; then, we evaluate the consequences of such scale dependence in the bispectrum and in $\fnl$. We must also take into account that the presence of statistical anisotropy is characterized by a privileged direction $\nn$ which, in turn, introduces a nontrivial structure in the higher order correlators that depends on the relative orientations of the wave vectors with respect to $\nn$. These relative orientations introduce a series of terms which modify and modulate the form of the higher order correlators and the levels of non-Gaussianity associated with them. We expect these terms and structures to encode potentially useful information about the presence of vector fields during inflation and to possibly leave signatures which hopefully could be significant from the observational point of view. If observed, these signatures would certainly give us clues about the role of vector fields during inflation. Quite to the contrary, the lack of these signatures would allow us to constrain and, eventually, rule out the vector fields as sources of the primordial curvature perturbation and as relevant pieces for the inflationary mechanism. 

In this paper, we have focused mainly on the relation between the $\fnl$ NG parameter (and the bispectrum) and the level of statistical anisotropy $\gz$ and its running with the scale $k$. To enter into the quantitative details of this relation, we have given expressions for the scale and shape dependence of $\fnl$ due to a scale dependent $\gz$ and evaluated the departure from the isotropic case through Eq. (\ref{ftofiso}). We have defined and evaluated the spectral indices in Eqs. (\ref{spectral-k}) and (\ref{spectral-alpha}) which measure the deviations from the scale and shape invariance in $\fnl$. These evaluations are the main result of this article. Measuring the impact of the statistical anisotropy in the bispectrum is a nontrivial task due to the richer structure and dependence on the configuration of the wave vectors in Fourier space, which indeed leads to several dependences that we should consider. In this regard, we have been rather conservative in our calculations and remained close to the current observational limits for some of the most relevant parameters, namely, $\gz$ and the spectral index of the primordial curvature perturbation $n_{\zeta}$. We have worked in detail on the massless vector field case and, as an extra consideration, we have set the vector perturbations to be scale invariant, which is a condition easy to reach in general inflationary vector field models. This condition is, however, not essential, the only necessary assumption being that $n_{\zeta} \approx 0.96$, as suggested by observations. As a result, within these limits, we have obtained that the departures from the isotropic case are strongly suppressed by $\gz$ and by the slow-roll parameters present in the spectral indices of the perturbations. The latter strongly suggests that a small level of statistical anisotropy  in the power  spectrum implies a small level of statistical anisotropy in the bispectrum and in higher order correlators as well. We need to go a bit farther from statistical isotropy, for instance, around $\gz \approx -0.1$, and for certain configurations, to obtain a significant effect in the bispectrum and in $\fnl$. Certainly, if those effects were not detected, the statistical anisotropy would be $|\gz| \lesssim 0.01$ or even smaller, meaning that there is no evidence for the breaking of rotational symmetry or, at least, not in the PS and the BS. 

To conclude, we would like to emphasize and recall that the statistical anisotropy may have its origin in the inflationary vector fields which introduce a rich and nontrivial structure in the correlators of the primordial curvature perturbation. We tried, in this paper, to go a step further in the direction of a precise characterization of the effects related to the statistical anisotropy by studying their impact on the correlation functions. It remains to be seen if the effects discussed here are relevant for modern high precision cosmology and if there is some possibility to perform measurements oriented to the detection of signatures of statistical anisotropy in the correlators associated with inflationary vector fields.

\section*{Acknowledgments}
This work was supported by COLCIENCIAS grant number 110656933958 RC 0384-2013 and by COLCIENCIAS -- ECOS NORD 
grant number RC 0899-2012 with the help of ICETEX.  J.P.B.A. and Y.R. were supported by VCTI (UAN) grant number 20131041. Y.R. was supported  by DIEF de Ciencias (UIS) grant number 5709. C.A.V.-T. acknowledges financial 
support  from Vicerrector\'ia de Investigaciones (Univalle) grant number 7924. J.P.B.A.  thanks Universidad del Valle for its warm 
hospitality and stimulating academic atmosphere during several stages of this project. 



\bibliographystyle{apsrev4-1}
\bibliography{Biblio}

\end{document}